\documentclass[sn-basic]{sn-jnl}% Basic Springer Nature Reference Style/Chemistry Reference Style

\usepackage{float}
\usepackage{graphicx}
\usepackage{dcolumn}% Align table columns on decimal point
\usepackage{bm}
\usepackage[utf8]{inputenc}
\usepackage[T1]{fontenc}
\usepackage{mathptmx}
\usepackage{etoolbox}
\usepackage{multirow}%
\usepackage{amsmath,amssymb,amsfonts}%
\usepackage{amsthm}%
\usepackage{mathrsfs}%
\usepackage[title]{appendix}%
\usepackage{xcolor}%
\usepackage{textcomp}%
\usepackage{manyfoot}%
\usepackage{booktabs}%
\usepackage{algorithm}%
\usepackage{algorithmicx}%
\usepackage{algpseudocode}%
\usepackage{listings}%

\begin{document}

%\preprint{}

\title[The evolutionary advantage of guilt]{The evolutionary advantage of guilt: co-evolution of social and non-social guilt in structured populations}


\author*[1]{\fnm{Theodor} \sur{Cimpeanu}}\email{theodor.cimpeanu@stir.ac.uk}

\author[2]{\fnm{Lu\'{i}s Moniz} \sur{Pereira}}\email{lmp@fct.unl.pt}

\author*[3]{\fnm{The Anh} \sur{Han}}\email{T.Han@tees.ac.uk}

\affil*[1]{\orgdiv{Biological and Environmental Sciences}, \orgname{University of Stirling}, \orgaddress{\city{Stirling}, \postcode{FK9 4LA}, \country{United Kingdom}}}

\affil[2]{\orgdiv{Department of Computer Science}, \orgname{Universidade Nova de Lisboa}, \orgaddress{\city{Caparica}, \postcode{2829-516}, \country{Portugal}}}

\affil[3]{\orgdiv{School Computing, Engineering and Digital Technologies}, \orgname{Teesside University}, \orgaddress{\city{Middlesbrough}, \postcode{TS1 3BX}, \country{United Kingdom}}}
% \author{Theodor Cimpeanu}
% \thanks{T.C. is supported by the John Templeton Foundation (grant no. 62281).}
% \email{tic1@st-andrews.ac.uk}
% \affiliation{School of Mathematics and Statistics, University of St Andrews, St Andrews KY16 9SS}

% \author{Lu\'{i}s Moniz Pereira}
% \email{lmp@fct.unl.pt}
% \affiliation{Department of Computer Science, Universidade Nova de Lisboa, 2829-516 Caparica, Portugal}

% \author{The Anh Han}
% \email{T.Han@tees.ac.uk}
% \affiliation{School Computing, Engineering and Digital Technologies, Teesside University, Middlesbrough TS1 3BX}

\newcommand{\ty}{\textcolor{yellow}}

\newcommand{\tr}{\textcolor{red}}
\newcommand{\tb}{\textcolor{blue}}
\newcommand{\tg}{\textcolor[rgb]{0.60,0.20,0.80}}

\abstract{
Building ethical machines may involve bestowing upon them the emotional capacity to self-evaluate and repent on their actions. While apologies represent potential strategic interactions, the explicit evolution of guilt as a behavioural trait remains poorly understood. Our study delves into the co-evolution of two forms of emotional guilt: social guilt entails a cost, requiring agents to exert efforts to understand others' internal states and behaviours; and non-social guilt, which only involves awareness of one’s own state, incurs no social cost. Resorting to methods from evolutionary game theory, we study analytically, and through extensive numerical and agent-based simulations, whether and how guilt can evolve and deploy, depending on the underlying structure of the systems of agents. Our findings reveal that in lattice and scale-free networks, strategies favouring emotional guilt dominate a broader range of guilt and social costs compared to non-structured well-mixed populations, so leading to higher levels of cooperation. In structured populations, both social and non-social guilt can thrive through clustering with emotionally inclined strategies, thereby providing protection against exploiters, particularly for less costly non-social strategies. These insights shed light on the complex interplay of guilt and cooperation, enhancing our understanding of ethical artificial intelligence.
}


\maketitle

% \begin{quotation}
% Imagine machines capable of not just acting according to their learned or programmed behaviour but by ''feeling guilt'', much like humans do. We explore the emotional landscape of ethical machines, bestowing upon them the capacity to self-evaluate and repent on their actions. We dissect guilt, often seen as a deeply human emotion, into two forms: social, where machines understand and react to others' emotions and actions, and non-social, solely focused on an awareness of internal states. Through complex simulations mimicking different social structures, we uncover surprising outcomes. Given certain networks, we unearth the protective power of guilt, offering defense against exploitation by way of clustering with other emotionally-prone individuals. These findings not only shed light on the intricate dance between social and non-social guilt in cooperative settings, but also hold profound implications for designing ethical and self-regulating artificial intelligence systems. 
% \end{quotation}

\section{Introduction}
"We are guilty for no reason, or just because we exist anyway, and are imperfect." - Peter J. Conradi \citep{Conradi2010} \\ \\
Machine ethics, focused on the potential for artificial intelligence (AI) to engage in moral conduct, represents an interdisciplinary open project for scientists and engineers \citep{frank:1998bv,pereira2021employing,Awad2018,kobis2021}.
An essential challenge within this field is the development of effective ways to represent emotions, like guilt, which are thought to shape human moral behavior, in computational models \citep{marsella2014computationally,gaudou2014moral,pereira2016programming,kowalczuk2016computational,Luis2017AAMAS,man2019homeostasis}.
Upon introspection, interpersonal guilt is present as a feeling of being worthy of blame for a moral offence committed against others. Carrying the burden of guilt, a person may subsequently work towards restoring an internal state untainted by blame, ensuring the absence of this distressing emotion \citep{tangney201323}.
The popular trend in research is to consider guilt more than shame as leading to reparative actions. This has been looked at in \citep{de2010restore}, stating that guilt entails reparative action when there is a conscious admission and accountability of the wrongdoing by the transgressor. 

Sociocentric and egocentric cultures supposedly have different emotional expressions and experiences of shame and guilt. Sociocentric cultures, which are more social-looking, tend to generate more of a sense of character-intrinsic general shame, while  more individualistic egocentric cultures  lead to a sense of specific action-intrinsic guilt in the transgressor \citep{Mayer2021}.
While shame and guilt are commonly thought to be synonymous, shame is defined as a self-centric emotion prompting the desire to hide and escape, whereas guilt is characterized by the motivation to engage in reparative actions \citep{Bastin2016, Keng2017}.
One feels guilt for having told a lie, but one feels shame for being a liar \citep[p.102]{Joyce2006}. Guilt concerns transgressions; shame involves shortcomings. Guilt urges reparative action; shame encourages social withdrawal. Guilt requires that a person judges that punishment is deserved, and even self-inflicted, and, moreover, that guilty others deserve it too.

%INCLUDE: {Joyce2006} Richard Joyce, The Evolution of Morality. The MIT Press, Cambridge MA, USA

When norms are well-established, societal members accept them as mandatory, internalise and comply with them, and experience guilt or shame when violating them. When internal sanctions do not support compliance over extended periods of time, external sanctions may be necessary  \citep{Billingham2020}.

\cite{tomasello2016}  emphasised %(pp.64,72,74,75,108,110,111,127) 
that prior joint objective commitment, or even a subjective commitment, makes guilt feel deserved. Guilt is a process of socially normative self-regulation. Besides self-punishment, there is a sense of "I ought not to have done that" implied in feeling guilty, be it even because of prior commitment. There is a normative, not just a strategic force in guilt, conducive to repair damage, with a concern to maintain one's cooperative identity. The conviction that one should prioritize doing the right thing has the potential to supersede self-centred motivations, extending beyond mere strategic reputation management. Guilt doesn't necessarily stem from a breach of any conventions; it isn't about feeling remorse solely for nonconformity itself. Rather, guilt is selectively aimed at one's previous judgment of moral rightness: "I thought at the time it was the right thing to do but, now aware of the consequences, I no longer do so." The overt response to guilt is thus to make reparations for harm done. The fact that guilt is a judgment about one's previous judgment comes out clearly in the fact that humans quite often feel the need to display their guilt overtly, in everything from body postures to verbal apologies. This display may pre-empt punishment from others, and  may additionally be  seen as strategic because it shows solidarity with those who can judge them harshly, and indeed, that it is accepted the negative judgment is deserved and legitimate. Reflective endorsement and guilt, therefore, represent a new kind of social self-regulation, an internalised and reflective self-regulation comprising multiple levels of moral judgement. Violators of moral norms punish themselves through feelings of guilt. They take on the perspective and attitude of the group when judging what they themselves have done.

%INCLUDE: {tomasello2016} Michael Tomasello, A Natural History of Human Morality. Harvard U.P., Cambridge MA, USA, 2016.

Guilt, as an emotion, incorporates a cognitive component, involving the acknowledgment that the subject has, in some way, violated a norm (survivor guilt being an exception) \citep[p.196]{Joyce2008}. There's even the potential for feeling guilty about the absence of guilt—a meta-emotion \citep[p.404]{Prinz2008}. The anticipation of guilt can drive normative conformity, even in the absence of an expected retaliatory response. When we anticipate the wrath of others or our own guilt, this can defeat the temptation to engage in subsequent harmful behaviour. Guilt promotes cooperative behaviour by adding an emotional cost to defection \citep[p.141]{Prinz2010}. Reciprocity violations provoke one's anger and an other's  guilt \citep[p.380]{Prinz2008}. Self-directed anger plays a different role than guilt. People get angry at themselves when they behave stupidly, but guilt arises when people violate norms (or commitments), and is especially likely when one causes harm to others. In addition, we feel more guilty about harming members of the in-group than those of the out-group. This finding lead some authors to conclude that guilt is an emotion that arises especially when there is a threat of separation or exclusion \citep[p.133]{Prinz2010}. Moreover, it highlights the importance of population networking with respect to guilt.

%INCLUDE: {Joyce2008} Richard Joyce, Aversions, Sentiments, Moral Judgments, and Taboos. In: Walter Sinnot-Armstrong (ed.), Moral Psychology Vol.1, The MIT Press, Cambridge MA, USA, 2008.

%AND: {Prinz2008} Jesse J. Prinz, Is Morality Innate? In: Walter Sinnot-Armstrong (ed.), Moral Psychology Vol.1, The MIT Press, Cambridge MA, USA, 2008.

%AND: {Prinz2010} Jesse J. Prinz and Shaun Nichols, Moral Emotions. In: John M. Doris (ed.), The Moral Psychology Handbook, Oxford U.P., UK, 2010.

Guilt-proneness has been highlighted as independent of anger; in other words, individuals with a predisposition to guilt are no more or less prone to anger than the general population \citep[pp.490-496]{tangney201323}. Yet, when confronted with anger, those with a propensity for guilt are more likely to channel it constructively, opting for non-hostile discussions, direct corrective actions, and a general aversion to aggression. Feeling guilty leads to positive intrapersonal and interpersonal processes. Expressions of guilt can strengthen relationships in a number of ways, especially in contexts requiring cooperation and interpersonal trust, based on assumptions of equity and fairness.

Guilt triggers self-debugging, as a result of an \textit{a posteriori} error detection in norm compliance. There is an expectation of correctness and a dissonance provoked by error, debugging being enabled by counterfactual reasoning \citep{Pereira2019,Pereira2024}. 

%INCLUDE: (Pereira2019) L. M. Pereira, F. C. Santos, Counterfactual Thinking in Cooperation Dynamics.  In: M. Fontaine et al. (eds.), Model-Based Reasoning in Science and Technology - Inferential Models for Logic Language, Cognition and Computation, pp. 69-82, SAPERE series, Springer, Cham, Switzerland, 2019. 

%AND: (Pereira2023) L. M. Pereira, F. C. Santos, A. Lopes, AI Modelling of Counterfactual Thinking for Judicial Reasoning and Governance of Law. In: H. S. Antunes, A. Oliveira (Eds.), Artificial Intelligence, Law and Beyond, Springer Nature AG, Cham, Switzerland, forthcoming 2023.

Righteousness is arguably the opposite of guilt \citep[p.17]{Daniel2003}. Guilt, arising from rule violation, contrasts with righteousness, a rewarding state achieved through rule adherence. When individuals "do the right thing," they experience a distinct positive emotion. Righteousness acts as a proxy for the rewards of conformity and serves to encourage it. Righteous individuals are willing to pay a price for resisting the temptation to swindle others.

%INCLUDE: {Daniel2003} Daniel M. T. Fessler and Kevin J. Haley, The Strategy of Affect: Emotions in Human Cooperation. In: Peter Hammerstein (ed.), Genetic and Cultural Evolution of Cooperation. The MIT Press, Cambridge MA, USA, 2003.

In social dilemmas games such as the Prisoners' Dilemma (PD) \citep{coombs1973reparameterization}, where  defection or cheating becomes the dominant strategy, defectors do better than cooperators regardless of whether their partners defect or cooperate \citep{sigmund:2010bo}. In such a situation, it is rational for both parties to defect, even though mutual defection is mostly overall worse than reciprocal cooperation. \cite{key:trivers1971} speculated that mutual evolution has promoted the emergence of guilt because it makes defection less attractive, with motivation from guilt becoming the dominant strategy due to attending social benefits. Individuals may gain materially by defecting, but guilt causes emotional suffering, and it is this suffering avoidance that encourages cooperation regardless of material gain.
\citep{Nesse2019} sustain that the temptation to defect arouses anxiety and defection arouses guilt, both aversive emotions that inhibit hasty selfishness. Guilt will motivate apologies, and/or self-punishment otherwise, and reparations are needed to re-establish trust.
Using an iterated PD game \citep{KetelaanAu2003} found that inducing guilt increased cooperativeness among previously uncooperative players.

%INCLUDE: T. Ketelaar and W. T. Au, The effects of feeling guilty on the behavior of uncooperative individuals in repeated social bargaining games: An affect-as-information interpretation of the role of emotion in social interaction. In: Cognition and Emotion 17:429-453 (2003).

From an evolutionary viewpoint, guilt is envisaged as an in-built mechanism that tends to prevent wrongdoing. Internal suffering and the need to alleviate it press an agent to their admission after wrongs are enacted, involving costly apology or penance, a change to correct behaviour, and an expectation of forgiveness to dispel the guilt-induced suffering.
The hypothesis then is that within a population, the emergence of guilt and its effects are evolutionary advantageous compared to a guilt-free population. Moreover, the magnitude of the advantage presumably depends on the population's actual network structure, since it governs who is in touch with whom and affected by whom \citep{Szabo2007,barabasi2014linked}, and determines the extent to which the social costs of guilt are globally worthwhile.

Inspired by the  discussed psychological and evolutionary studies of guilt and cooperation in networks \citep{santos:2008:nature,szabo:2007jt,barabasi2014linked,guo2023third,cimpeanu2023social, flores2024evolution, Xia2023, zhou2025}, here we provide a theoretical account of the evolution of costly guilt-prone behaviours in the context of distributed Multi-Agent Systems (MAS), with the overarching aim of achieving new insights for the design and engineering of cooperative, self-organised systems. 
Resorting to methods from Evolutionary Game Theory (EGT)  and agent-based simulations \citep{sigmund:2010bo,perc2017statistical}, we study the evolution of social vs non-social aware guilt in differently structured populations.

We shall examine  whether (non-)social guilt can evolve in such structured populations, e.g. through clustering of similarly emotionally prone individuals.
Social guilt, and social emotions in general, depend upon awareness of the thoughts, feelings or actions of others in  the environment \citep{hareli2008s,burnett2009development}. 
Thus, choosing to be social can be (much) more costly compared to being non-social, requiring efforts to understand or be more aware, through observation of others' thoughts and feelings and the context behind their actions; while  non-sociality only requires awareness of one's own internal physical state. Non-social guilt is an internal mechanism -- \textit{I} did something bad to another, therefore \textit{I} feel suffering as a consequence. This requires no awareness of the other's emotional state. Social guilt is more akin to a norm. I only feel guilt if the other person is also repentant. In other words, I only feel guilt if they also feel when they have broken a norm, so that there is something for me to feel guilty about.
Hence, one might inquire whether and when such a more cost-efficient  non-social strategy can evolve (though more easily exploitable as we will see), depending on the specific underlying network structure.

Herein, we fundamentally extend and generalise the work we set forth in \citep{Luis2017AAMAS}, which constructed theoretical models representing guilt to study its role in promoting pro-social behaviour, in the context of EGT using the Iterative Prisoners’ Dilemma (IPD). Guilt was modelled in terms of two joint features. Firstly, guilt involving a record of transgressions formalized as a counter tracking the number of offences. Secondly, guilt involving a threshold over which the guilty agent must alleviate its strained internal state, by means of deliberate change of behaviour plus self-punishment, as required by the negative feelings of guilt, changes such that would affect the game's payoff for the guilty party. 
 
% The remainder of the paper is structured as follows. We start off with related work, proceed to our models and methods, present their results, and terminate with concluding remarks. Moreover, we provide additional results as Supporting Information (\textbf{SI}).  

\section{Models and Methods}
First we recall the Iterated Prisoner's Dilemma (IPD) game and the definition of guilt-prone strategies, as described in \citep{Luis2017AAMAS}. 
Next, we describe our model where social and non-social guilt strategies are in co-presence. 
Then, the methods for analysing the model, namely  stochastic evolutionary dynamics in well-mixed populations,  and the agent-based simulations in networks, are in turn detailed. 

\subsection{Iterated Prisoners' Dilemma (IPD)}
In each round of the IPD, two players engage in an PD game interaction where its outcomes are defined by the following payoff
matrix (for the row player)
\[
 \bordermatrix{~ & C & D\cr
                  C & R & S \cr
                  D & T & P  \cr
                 }.
\]
A player who chooses to cooperate (C) with another who defects (D) receives the sucker's payoff $S$, whereas the defecting player gains the temptation to defect, $T$. Mutual cooperation (resp., defection) yields the reward $R$ (resp., punishment P) for both players.  
Depending on the value ordering of these four payoffs, different social dilemmas arise \citep{coombs1973reparameterization,sigmund:2010bo}. In this work we are namely concerned with the PD, where $T > R > P > S$. 
In a single round, it is always best to defect, because less risky, but cooperation may be rewarding if the game is iterated. In the IPD, it is also required that mutual cooperation be preferred over an equal probability of unilateral cooperation and defection (i.e. $2R > T +S$); since otherwise, alternating between cooperation and defection, would lead to a higher payoff than mutual cooperation. The PD is repeated for a number of rounds, $\Omega$.

For a convenient interpretation of results, we also consider the simplified version of the PD, the Donation game \citep{sigmund:2010bo}, where the payoff entries are specifically described via the cost $c$ ($c > 0$) and benefit $b$ ($b > c$) of cooperation, as follows: $T = b$, $R = b-c$, $P = 0$, $S = -c$.

%The repetition in the IPD is modeled as follows. After the current interaction, another interaction between the  interacting players occurs with probability   $\Omega \in (0,1)$, resulting in an average  of $(1-\Omega)^{-1}$ rounds in the game. 

%\ty{The IPD is played under the presence of noise, that is, an intended action, $C$ or $D$, can fail, and become its opposite, with probability $\alpha \in [0,1]$ \citep{key:Sigmund_selfishnes}.

\subsection{Guilt modelling in IPD}

%\tr{[Add an illustration of the guilt and PD game?]}

%\tr{[Add a table to summarize the strategies]}

We base our model and analysis on the approach set forth in 
 \citep{Luis2017AAMAS}, which formalizes guilt as an aspect of an agent’s genotypical strategies, and is quantified in terms of a threshold, G. In this model, $G \in [0, +\infty]$, and guilt at any given time is characterized by a transient level of guilt, $g$ ($g \geq 0$). As the experiment begins, $g$ for every agent is set to $0$. It increases by $1$ after each action that the agent considers  wrong. After several accumulated wrongdoings result in $g$ reaching that agent’s threshold of guilt, $g \geq G$, the agent can choose to (or not to) act to reduce its guilt level $g$ below that threshold. The model retains the mechanism of guilt alleviation described above, whereby guilt can be alleviated by apologising to offended partners, or by suffering guilt through self-punishment whenever apology to offended partners is not an option. In the sequel we will suppose the latter case. Either way, the guilty party suffers a cost.
 Indeed, the alleviation of guilt is costly, this cost being quantified in terms of $\gamma$ ($\gamma \geq 0$), whenever $g$ is decreased by $1$. In accordance with this definition, agents can be characterized with respect to different guilt thresholds. Some may be incapable of suffering guilty feelings, meaning their $G = +\infty$. Others may be extremely prone to guilt, suffering guilty feelings with any first mistake, so for them $G = 0$. These are the only two cases to be considered below.

\subsection{Social vs. non-social guilt in co-presence} 
In this setting, a strategy is described by three factors or components: \\
 \textbf{(I) Guilt threshold $G$.} Since we shall focus in the current work  on understanding the evolution of social guilt behaviours, or their absence, as well as the impact on them of network structures, we  consider two basic types of guilt thresholds
\begin{itemize}
 \item $G = +\infty$: In this type of agent the guilt level $g$ will never reach the threshold no mater how many times they defect; hence,  they  never need to reduce $g$, and consequently never pay the guilt cost $\gamma$. In other words, this type of agent experiences no guilt feeling. They are dubbed (guilt-)unemotional agents.  
 \item  $G = 0$: whenever this type of agent defects, it becomes immediately true that  $g > G$; hence,  the agent needs to act immediately to reduce $g$, by paying  $\gamma$. In other words, this type of agent always feels guilty subsequent to just a single a wrongdoing, viz. defection. They are dubbed {(guilt-)}emotional agents.  
\end{itemize} 

\noindent \textbf{(II) Decision making in the IPD}. An agent can choose to play either C or D in a PD and, given the agent's guilt threshold $G$, if its ongoing guilt level $g$ reaches  $G$, they can choose whether to change their behaviour from D to C (to avoid further emotional pain and cost). \\

\begin{figure*}[t]
    \centering
    \includegraphics[width=1\linewidth]{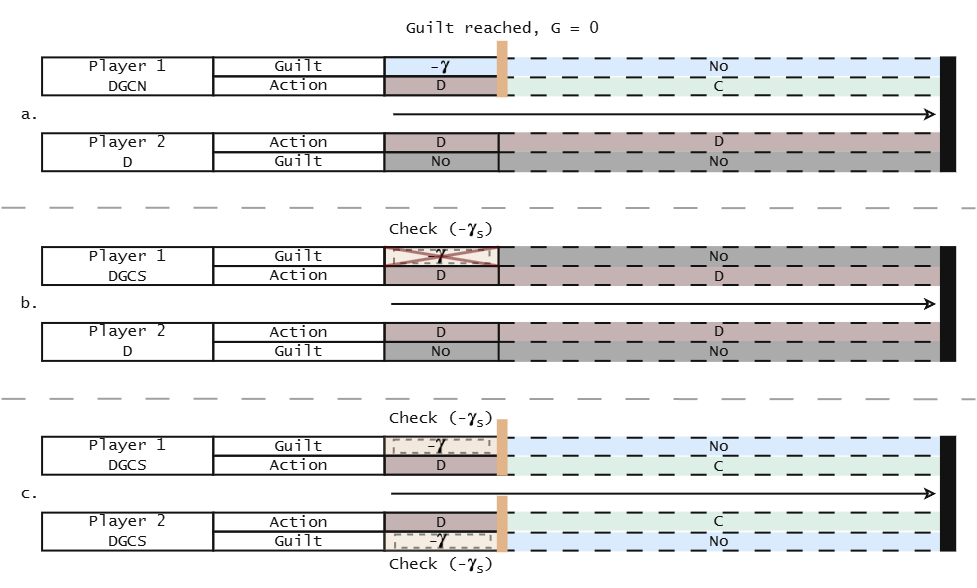}
    \caption{\textbf{Diagrams representing repeated interactions between emotionally-prone players.} In \textbf{a}, an emotionally adaptive non-social defector interacts with another defector; they feel guilty after one interaction (here $G = 0$) and change their behaviour to prevent further internal pain. In \textbf{b}, an emotionally adaptive social defector checks whether their partner felt guilt for their actions, and, if so, does not feel guilt nor change behaviour in future interactions. In \textbf{c}, two of the adaptive social players interact; after checking, they feel guilty for their first transgressions, and so cooperate in future interactions.}
    \label{fig:diagram}
\end{figure*}

\noindent \textbf{(III)  Sociability vs non-sociability about when to feel guilty}. 
The emotional agents can choose to be non-social  or social, regarding the way they express their emotion. 
To be social, agents need an extra effort such as signalling their guilt or observing the co-player's guilt, as this observation will affect how they themselves will react (see Figure \ref{fig:diagram} for a visual representation of interactions with social and non-social strategists).
Hence, we assume there is always an additional cost, $\gamma_s$, to being social. 

Overall, since we do not yet consider noise in IPD (i.e. non-deliberate mistakes)  in this work, there are in total six possible strategies \footnote{There can be other strategies such as emotional (i.e. $G = 0$) cooperators who always cooperate and thus never feel guilty. But as we are not modelling noise in this work, this strategy is equivalent to C in all interactions, and can consequently be removed from our analysis. }, denoted as follows 
\begin{enumerate}
    \item     Unemotional cooperator (C): always cooperates (C), unemotional (i.e. $G = +\infty$). Does not feel guilt, does not change behaviour.
    \item     Unemotional defector (D): always defects (D), unemotional (i.e. $G = +\infty$).  Does not feel guilt, does not change behaviour.
      \item    Emotional non-adaptive defector that is non-social (DGDN): always defects (D), feels guilty (G) after one wrongdoing (i.e. $G = 0$), does not change its behaviour (thereby the second D), regardless of what the co-player feels (hence its non-sociability N).
    \item   Emotional adaptive defector that is non-social (DGCN):  defects initially (D), feels guilty (G) after one wrongdoing (i.e. $G = 0$), changes its behaviour from D to C (hence the C), regardless of what the co-player feels (hence its non-sociability N).
     \item    Emotional non-adaptive defector that is social (DGDS): always defects (D), feels guilty (G) after one wrongdoing (i.e. $G = 0$) but only if their co-player also feels guilty after a wrongdoing (hence its sociability S), but does not change its behaviour (hence the second D). 
    \item   Emotional adaptive defector that is social (DGCS):  defects initially, feels guilty after one wrongdoing (i.e. $G = 0$) but only if their co-player  also feels guilty after a wrongdoing (hence its sociability S), and changes its behaviour from D to C (hence the C). 
\end{enumerate}
From the above costs, we can derive the  payoff matrix for these six strategies (for the row player), as follows
\begin{equation} 
\hspace*{-0.75cm}
 \label{payoff_matrix_6strats} 
\bordermatrix{~ & C & D &  DGDN & DGCN &  DGDS & DGCS \cr
                  1 &R & S &  S & \frac{S+R \Theta}{\Omega } &  S & \frac{S+R \Theta}{\Omega }   \cr
                  2 & T & P &  P & \frac{P+T \Theta}{\Omega } &  P & P \cr
                  3 & T-\gamma  & P - \gamma &  P - \gamma  & \frac{P+T \Theta}{\Omega } - \gamma & P - \gamma & \frac{P+T \Theta}{\Omega } - \gamma \cr
                  4  &  \frac{T-\gamma +R \Theta}{\Omega } &\frac{P-\gamma +S \Theta}{\Omega } &  \frac{P-\gamma +S \Theta}{\Omega } & \frac{P-\gamma +R \Theta}{\Omega }  & \frac{P-\gamma +S \Theta}{\Omega }   & \frac{P-\gamma  +R \Theta}{\Omega } \cr
                5 & T-\gamma- \gamma_s  & P- \gamma_s & P- \gamma - \gamma_s    & \frac{P +T \Theta}{\Omega } -\gamma- \gamma_s& P- \gamma- \gamma_s & \frac{P+T \Theta}{\Omega }  -\gamma- \gamma_s  \cr
                  6  &  \frac{T-\gamma- \gamma_s +R \Theta}{\Omega } & P- \gamma_s & \frac{P - \gamma - \gamma_s +S \Theta}{\Omega }  & \frac{P-\gamma- \gamma_s +R \Theta}{\Omega } &  \frac{P-\gamma - \gamma_s+S \Theta}{\Omega } & \frac{P-\gamma- \gamma_s +R \Theta}{\Omega } \cr
                  },
\end{equation}
 where we employ $\Theta = \Omega -1$ just for the purpose of neater representation.
%We can observe that DGCN is always risk-dominant against DGCS when $\gamma_s > 0$. 

In order to understand when guilt can emerge and promote cooperation, our EGT modelling study below analyses  whether and when emotional strategies, i.e. those with $G = 0$, can actually overcome the disadvantage of the incurred costs or fitness reduction  associated with the guilt feeling and its alleviation,  and as a consequence be able to disseminate throughout the population.

Previous work shows that some emotional guilt-based responses only make sense when the  co-player is not attempting to harm you too, or else attempting to harm you but feeling guilty as well \citep{Luis2017AAMAS}. That is, guilt needs to be social to  prevail in social dynamics. The main reason is that players who feel guilty after a wrongdoing, regardless of others' behaviours (i.e. whether these others signal guilt or else are observed to feel guilty), would be exploited by non-emotional defectors (i.e. the D strategy).  We argue that, since being social is costly, since agents need to observe and understand others' actions and feelings, non-social guilt might conceivably be more cost-efficient and prevail in network environments where they might be protected from such D strategy exploiters by non-connection.  
Because previous guilt modelling work only looked at well-mixed networked populations, wherein all individuals in the population interact with one another, it was not possible to consider such eventual network connectivity protection. 
To bridge this gap, in this work we shall address several cases of structured populations, wherein players interact only with their direct neighbours.

%To that purpose, we analyse three different models, which differ in the way guilt influences the preferences of  the focal agents, where the preferences are determined by the payoffs in the matrices. 

\subsection{Evolutionary Dynamics in Well-Mixed Populations}
In our analysis, individuals'  payoffs signify their \emph{fitness} or social \emph{success}, and  evolutionary dynamics is shaped  by social learning \citep{key:Hofbauer1998,key:Sigmund_selfishnes}.  In this process, the agents who achieve higher success are more likely to be imitated by their others. In the current work, social learning is characterised by  the so-called pairwise comparison rule \citep{traulsen2006}, a common approach in Evolutionary Game Theory (EGT).
This rule assumes that an agent $A$ with a fitness value $f_A$ adopts the strategy of another agent $B$ with a fitness value $f_B$ with a probability $p$ determined by the Fermi function:
\begin{equation}
\label{eq:Fermi}
p_{A, B}=\left(1 + e^{-\beta(f_B-f_A)}\right)^{-1}.
\end{equation}
The parameter $\beta$ denotes the 'imitation strength' or 'intensity of selection,' signifying how strongly agents base their decision to imitate on the fitness difference between themselves and their opponents.  When $\beta=0$, the system reaches the limit of neutral drift, where imitation decisions are entirely random. As $\beta$ increases, imitation becomes increasingly  more deterministic.
 Consistent with previous works and human behavioural experiments \citep{Szabo2007,zisisSciRep2015,randUltimatum}, we adopt $\beta = 1.0$ in the main text, which also allows us to compare directly with the previous guilt model of \citep{Luis2017AAMAS}.
 
% {Despite using small mutation, it's been shown that the results are robust for larger mutation rates. Also, the analytical conditions do not depend on the small mutation limit assumption. }

\begin{figure*}[h]
    \centering
    \includegraphics[width=0.8\linewidth]{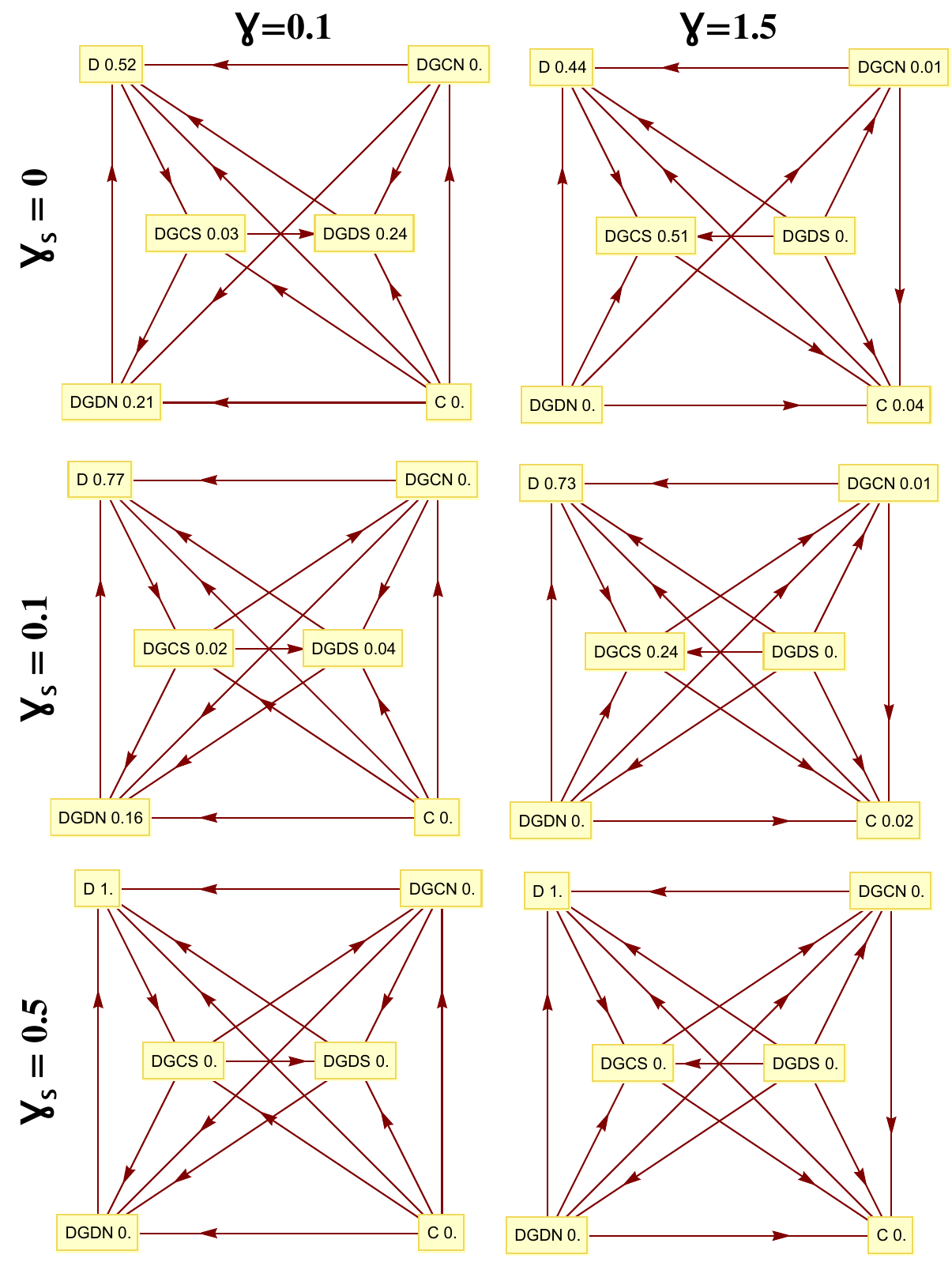}
    \caption{\textbf{Markov diagrams and stationary distributions (well-mixed populations).} Transitions direction among strategies, where the arrows show the direction where the transition probability is stronger than the reverse.  The results are in line with risk-dominance analysis (in Section III-A).   Other parameters: $N = 100$, $\Omega = 10$,  $R = 1$,  $S = -1$, $T = 2$, $P = 0$.  }
    \label{fig:markov}
\end{figure*}
 
In the absence of mutations or strategy exploration, the end states of evolution are inevitably monomorphic. Once such a state is attained, it cannot be escaped through imitation. To account for this, we introduce the assumption that, with a certain mutation probability,  an agent may switch randomly to a different strategy without the necessity of imitating another agent.  In the limit of small mutation rates, it is established that the dynamics will proceed with, at most, two strategies in the population. As a result, the behavioural dynamics can be succinctly delineated through a Markov Chain, where each state represents a monomorphic population. In this process,  the transition probabilities are given by the so-called fixation probability of a single mutant \citep{key:imhof2005,key:novaknature2004}. The resulting Markov Chain has a stationary distribution, which characterises the average time the population spends in each of these monomorphic end states (see  some examples in Figure \ref{fig:markov}).

Consider a population of size $N$. Denote $\pi_{X,Y}$ the payoff a strategist $X$ obtains in a pairwise interaction with strategist $Y$ (as defined by the payoff matrices). Assuming at most two strategies in the population, with $k$ agents using strategy A ($0 \leq k \leq N$)  and $(N-k)$ agents using strategy B. Thus, the (average) payoff of the agent that uses  A (similarly for B) is 
\begin{equation} 
\label{eq:PayoffA}
\begin{split} 
\Pi_A(k) &=\frac{(k-1)\pi_{A,A} + (N-k)\pi_{A,B}}{N-1}.%,\\
%\Pi_B(k) &=\frac{k\pi_{B,A} + (N-k-1)\pi_{B,B}}{N-1}.
\end{split}
\end{equation}

The probability to change the number $k$ of agents using strategy A by $\pm$ $1$ in each time step is given  as \citep{traulsen2006} 
\begin{equation} 
T^{\pm}(k) = \frac{N-k}{N} \frac{k}{N} \left[1 + e^{\mp\beta[\Pi_A(k) - \Pi_B(k)]}\right]^{-1}.
\end{equation}
Now, the fixation probability of a single mutant with a strategy A in a population of $(N-1)$ agents using B can be written as \citep{traulsen2006,key:novaknature2004}
\begin{equation} 
\label{eq:fixprob} 
\rho_{B,A} = \left(1 + \sum_{i = 1}^{N-1} \prod_{j = 1}^i \frac{T^-(j)}{T^+(j)}\right)^{-1}.
\end{equation} 
%In the limit of neutral selection (i.e. $\beta = 0$), $\rho_{B,A}$ equals the  inverse of population size, $1/N$. 

Considering a set  $\{1,...,q\}$ of distinct strategies, these fixation probabilities establish a transition matrix $M = \{T_{ij}\}_{i,j = 1}^q$ of a Markov Chain. Here,  $T_{ij, j \neq i} = \rho_{ji}/(q-1)$ and  $T_{ii} = 1 - \sum^{q}_{j = 1, j \neq i} T_{ij}$. The normalized eigenvector associated with the eigenvalue 1 of the transposed matrix of $M$ yields the stationary distribution described above \citep{key:imhof2005}, depicting the relative time the whole population spends adhering to each of the strategies.

\subsubsection*{Risk-dominance} A crucial approach to comparing  two strategies A and B is to ascertain the direction in which the transition is stronger or more probable---whether an A mutant fixates in a population of agents using B ($\rho_{B,A}$) or a B mutant fixates in the population of agents using A ($\rho_{A,B}$). It can be demonstrated that the former is stronger in the limit of large $N$, if the following (risk-dominant) condition is satisfied \citep{key:novaknature2004,key:Sigmund_selfishnes}, if 
%\begin{equation} 
%(N-2)\pi_{A,A} + N\pi_{A,B} > (N-2)\pi_{B,A} + N\pi_{B,B}
%\end{equation} 
%which,is simplified to 
\begin{equation} 
\label{eq:riskdom}
\pi_{A,A} + \pi_{A,B} > \pi_{B,A} +  \pi_{B,B}.
\end{equation}

\subsection{Agent-based Simulations and Network Structures}
\subsubsection{Network Topologies}
Connections within a network not only signify proximity in terms of interaction (indicating with whom the agents can interact) but also in an observational sense (highlighting whom the agents can imitate). Thus, the network of interactions aligns with the imitation network \citep{ohtsuki2007breaking}. As each network type converges at different rates and naturally presents various degrees of heterogeneity, we employ varying population sizes in our  experiments to investigate this, while optimising run-time.
 
Well-mixed populations, where all interact with all, provides a suitable baseline scenario, as  no specific heterogeneous interaction structure is present. Considering the realm of structured populations, we take a step further, probing the role of network properties and structural heterogeneity in cultivating the evolution of guilt-prone behaviours. Initially, we examine square lattice (SL) populations of size $N = 30 \times 30$, employing periodic boundary conditions, a  widely adopted population structure in population dynamics and evolutionary games (for a survey, see \citep{Szabo2007}), wherein each agent can only interact with its four immediate neighbours. While the SL  introduces a network structure, it is noteworthy that all nodes within this setup can be conceptualized as structurally equivalent. 

Taking our investigation a step beyond, we explore complex networks in which the network portrays a heterogeneity that mirrors the power-law distribution of wealth (and opportunity) characteristic of real-world settings. The Barab{\'a}si and Albert (BA) model \citep{barabasi1999emergence} is one of the most widely adopted models used in the study of such heterogeneous, complex networks. Key features of the BA model include adherence to a \textit{preferential attachment} rule, a low clustering coefficient, and a characteristic \textit{power-law degree distribution}. To elucidate the concept of preferential attachment, we outline below the construction process of a BA network.

Starting from a small set of $m_0$ interconnected nodes, each new node selects and establishes a link with $m$ older nodes following a probability proportional to their degree (the number of its edges).  This process continues until the network reaches the desired size of  $N$. This will produce a network characterised by a power-law distribution, $p_k \sim k^{-\chi}$, where the exponent $\chi$ is its $\textit{degree exponent}$ \citep{Barabasi2016}. Notably, the network has high degree correlation among nodes, featuring a skewed degree distribution with a prolonged tail. A few hubs in the network  attract an increasing number of new nodes, which attach as the network grows (in a typical \textit{``rich-get-richer''} scenario). The power-law distribution observed in BA networks mirrors the heterogeneity found in various real-world networks. The average connectivity of the obtained scale-free network is $z = 2m$. For all of our experiments, we pre-seed 10 different SF networks of size $N = 1000$ and an average connectivity of $z = 4$, in alignment with the number of neighbours in a square lattice.

\begin{table}%[h]
\centering
\begingroup
\renewcommand{\arraystretch}{1.5}
\begin{tabular}{lc}
\hline
 Parameter  &  Symbol %& Range/Value Analysed
 \\ 
\hline
 Population size & $N$ %& 100 
 \\
  Cost of cooperation & $c$ %& 100 
 \\
 Benefit of cooperation & $b$ %& 100 
 \\
 Intensity of selection & $\beta$  %& \{$0.1$\}  
 \\
% Payoff matrix  & R, S, T, P  & \{$1, -1, 2, 0$\} \\
 Guilt cost & $\gamma$  %& $[0,6]$  
 \\
 Social cost of guilt  & $\gamma_s$  %& $[0,1]$  
 \\
 Number of rounds in IPD & $\Omega$  %&  $[0,6]$   
 \\
Guilt threshold  & $G$ \\ %& $[0,0.2]$ 
\end{tabular}\
\endgroup
 \caption{Model parameters.}
 \label{table:parameters}
\end{table}

\subsubsection{Computer Simulations}
Initially each agent is designated as one of the six strategies (i.e., C, D, DGDN, DGCN, DGDS, DGCS), with equal probability. At each time step, each agent plays the PD with its immediate neighbours. The fitness score for each agent is the sum of the payoffs in these encounters. At the end of each step, an agent $A$ with fitness $f_A$ chooses to copy the strategy of a randomly selected neighbour agent $B$ with score $f_B$, with a probability given by the Fermi function \citep{Szabo2007} in Equation \ref{eq:Fermi}. Similar to the well-mixed setting above,  we set $\beta = 1$ in our simulations. %With a given probability $\mu$, this process is replaced instead by a randomly occurring mutation. 

We simulate this evolutionary process until a stationary state or a cyclic pattern is reached. %Similarly to \citep{nowak1992evolutionary}, all the simulations in this work (described in next sections) converge quickly to such a state. 
For the sake of a clear and fair comparison, all simulations are run for $10^6$ steps. %As we use an asymmetric update approach (only one individual is eligible for imitation at each time step), the number of update steps in one generation coincides with the size of the network.
Moreover, for each simulation, the results are averaged over the final $10^5$ generations, in order to account for the fluctuations characteristic of these stable states. 
Furthermore, to improve accuracy, for each set of parameter values, the final results are obtained by averaging 30 independent realisations (20 for scale-free networks due to computational overheads and the additional pre-seeding of networks, viz. 200 replicates for SF networks).

\section{Results}
Given the model and methods described above (see Table \ref{table:parameters} for a summary of the parameters), we first derive analytical conditions for when guilt-prone strategies can be viable and promote the evolution of enhanced cooperation. 
Next, we  obtain simulated numerical results for the well-mixed population setting, validating the analytical conditions.  
We then show results from our extensive agent-based simulations in structured population settings. 

\subsection{Risk dominance  of guilt-prone strategies}
To start with, we obtain analytical conditions for when  guilt-prone  strategies can be  evolutionarily viable against other strategies. For that, we apply the risk-dominance criteria in \ref{eq:riskdom} to the payoff matrix given in \ref{payoff_matrix_6strats}.

First, DGCS is risk-dominant against DGDS if 
\begin{equation}
    \gamma + \gamma_s > \frac{T-R+P-S}{2} = c.
    \end{equation}
The condition for DGCS to be risk-dominant against C is the reverse of that of against DGDS above.
DGCS is risk-dominant against DGDN if 
\begin{equation}
    (\Omega-1)\gamma - \gamma_s > (\Omega-1)\frac{T-R+P-S}{2} = (\Omega-1)c.
    \end{equation}
 It can be seen that this condition subsumes the one for risk-dominance against DGDS above. 
 Also, for this inequality to hold the necessary condition is $\gamma > c$.

Now, DGCS is risk-dominant against  D if
\begin{equation}
\label{equation:riskDGCSvsD}
\gamma + (\Omega+1)\gamma_s < (\Omega-1)(R-P) =  (\Omega-1)(b-c).
  \end{equation}
DGCS is risk-dominated by  DGCN whenever $\gamma_s > 0$. They are neutral when $\gamma_s = 0$.
However, DGCN is always risk-dominated by D. 
Thus, there is a cyclic pattern from DGCS (social guilt), to DGCN (non-social guilt), to D (non-emotional defectors), and back to DGCS, whenever the condition  in Equation \ref{equation:riskDGCSvsD} holds. That  occurs when  $\gamma$ and $\gamma_s$ are sufficiently small.  Fixing $c$, the latter condition is more easily satisfied for a more beneficial PD (i.e. large $b$). 

Moreover, for  DGCS to be risk-dominant against all the defective strategies (i.e. all but C and DGCN), the guilt cost $\gamma$ needs to be sufficiently large; that is, at least the cost of cooperation, $c$. Given that, the smaller the social cost, the easier it is for these conditions to be satisfied. The upper bound of this cost is $\frac{(\Omega-1)(b-c)}{\Omega+1}$.

\subsection{Well-mixed populations: Evolution of social guilt  and the eradication of non-social guilt}
To illustrate the above obtained analytical observations, Figure \ref{fig:markov} shows the stationary distribution and transition directions in a well-mixed population of the six strategies (see Methods).
We can see that the directions of transition, showing risk-dominance of the strategy at the end of the transition or arrow, corroborate the analytical conditions. We further perform replicator dynamics and a brief analysis of equilibrium points in Sections 1.3 and 1.4 of the supplementary material (see also Figures S4 and S5).  

\begin{figure*}[h]
    \centering
    \includegraphics[width=\linewidth]{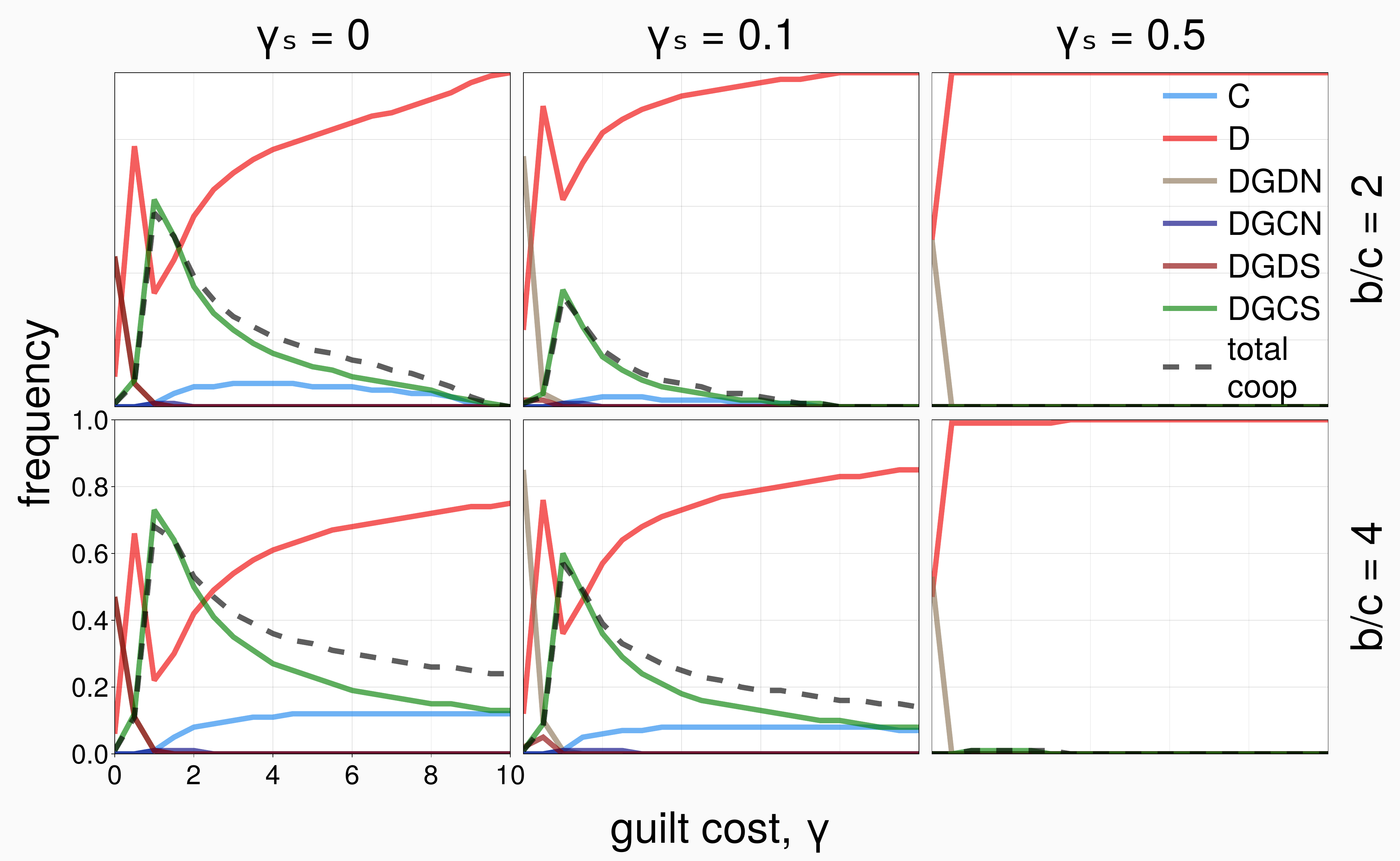}
    \caption{Strategies' frequency and  total cooperation level as a function of the guilt cost, $\gamma$ (well-mixed, $N = 100$, $\Omega = 10$). }
    \label{fig:dependgamma}
\end{figure*}

Figure \ref{fig:dependgamma} shows the long-term frequencies of the strategies and the total level of cooperation in the population,  for varying the guilt cost $\gamma$, for different benefits $b = 2$ (first row) and $b = 4$ (second row), and for different social costs $\gamma_s$. 
We  observe that, when the social cost $\gamma_s$ is sufficiently small, there is an intermediate value of the guilt cost $\gamma$ (around $\gamma = c$), which leads to an optimal frequency of DGCS and the total cooperation in the population. 
When $\gamma$ is too small, DGCS is dominated by DGDN (and DGDS) (see also Figure \ref{fig:markov}, first column).  
When $\gamma$ is larger, D frequency increases and dominates the population, despite being still dominated by DGCS (see Figure \ref{fig:markov}, second column). There is now a transition from DGCS to C which is strongly dominated by D.  Comparing the first and second rows of Figure \ref{fig:dependgamma}, a  higher level of cooperation is achieved for a larger benefit of cooperation $b$.

In short, we can observe that social guilt (DGCS) can evolve in the well-mixed population setting when the social cost is sufficiently small, reaching its peak around $\gamma \approx c$. Non-social guilt does not evolve at all in this setting, even when it dominates DGCS (whenever $\gamma_s > 0$, see Figure \ref{fig:markov}, second and third rows), as DGCN is always strongly dominated by D.  

\begin{figure*}[h]
    \centering
    \includegraphics[width=\linewidth]{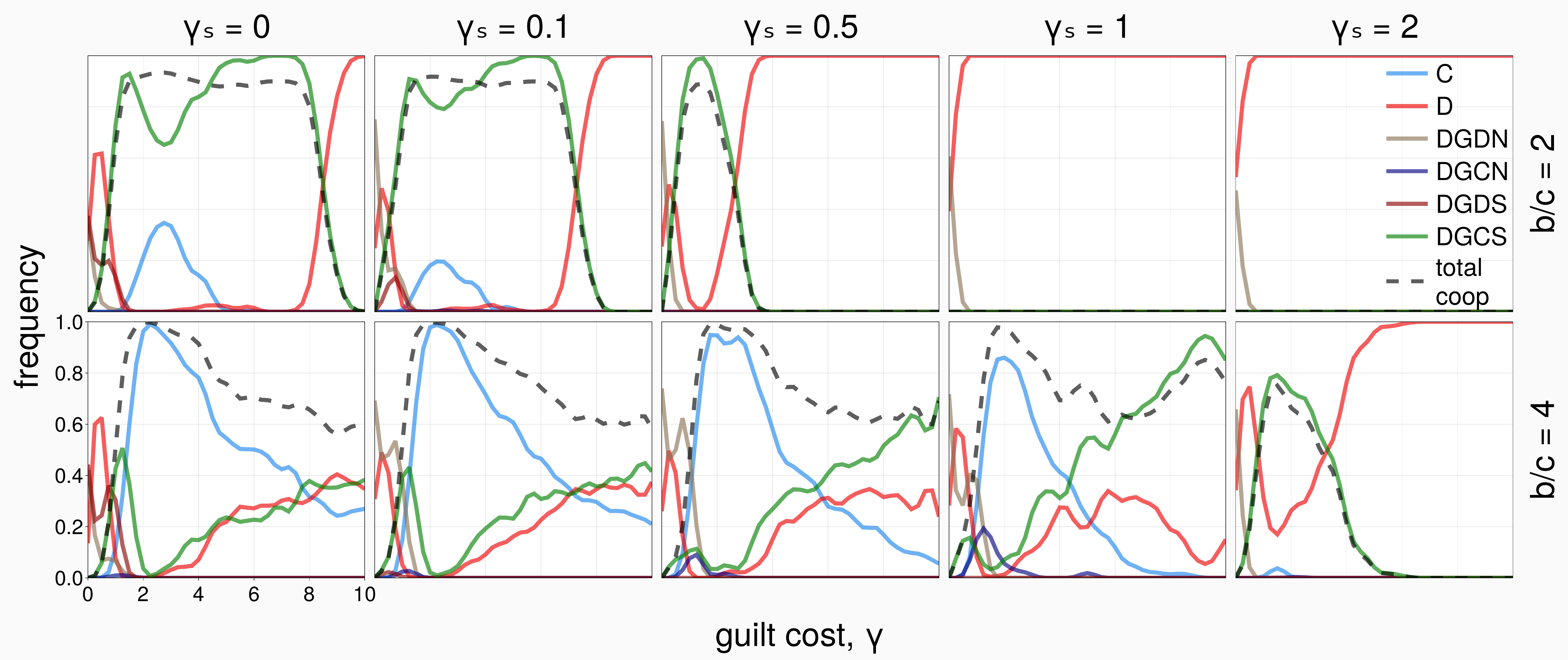}
    \caption{Strategies' frequency and the total cooperation level as a function of the guilt cost, $\gamma$ (square lattice,  $N = 900$, $\Omega = 10$).}
    \label{fig:gamma_lattice}
\end{figure*}

\subsection{Structured populations enhance social guilt and enable the emergence of non-social guilt}

We study the effect of spatial or structured populations on the evolutionary dynamics and outcomes of guilt-prone strategies (both social and non-social), as well as cooperation. 
Firstly, we consider results in the square lattice (SL) network, a regular (homogeneous) structure, see Figure \ref{fig:gamma_lattice}. 
We observe that, for a small benefit of cooperation $b = 2$ (top row), for sufficiently small social costs $\gamma_s$ (0 and 0.1), DGCS dominates the population over a wide range of $\gamma$, between approximately $1 < \gamma < 8$. Interestingly, there is also a chance for C to emerge. Moreover, when $b$ is larger (bottom row), C even dominates the population for a wide range of $\gamma$ and $\gamma_s$. DGCS dominates when $\gamma$ is sufficiently high. Interestingly, in such networked populations, even non-social guilt strategies can survive with some frequency when the social cost is non-negligible, see $\gamma_s = 0.1, \ 0.5$ and $1$ at intermediate ranges of $\gamma$. Overall, we observe significantly higher levels of cooperation and guilt-prone strategies for a wider range of both guilt and social costs, compared to well-mixed populations.

%In both lattice and SF we can see DGC players and cooperation are dominant for a larger range of the guilt cost $\gamma$, compared to well-mixed population settings, see Figures \ref{fig:gamma_lattice} for lattice and \ref{fig:gamma_sf} for SF networks.

\begin{figure*}[h]
    \centering
    \includegraphics[width=\linewidth]{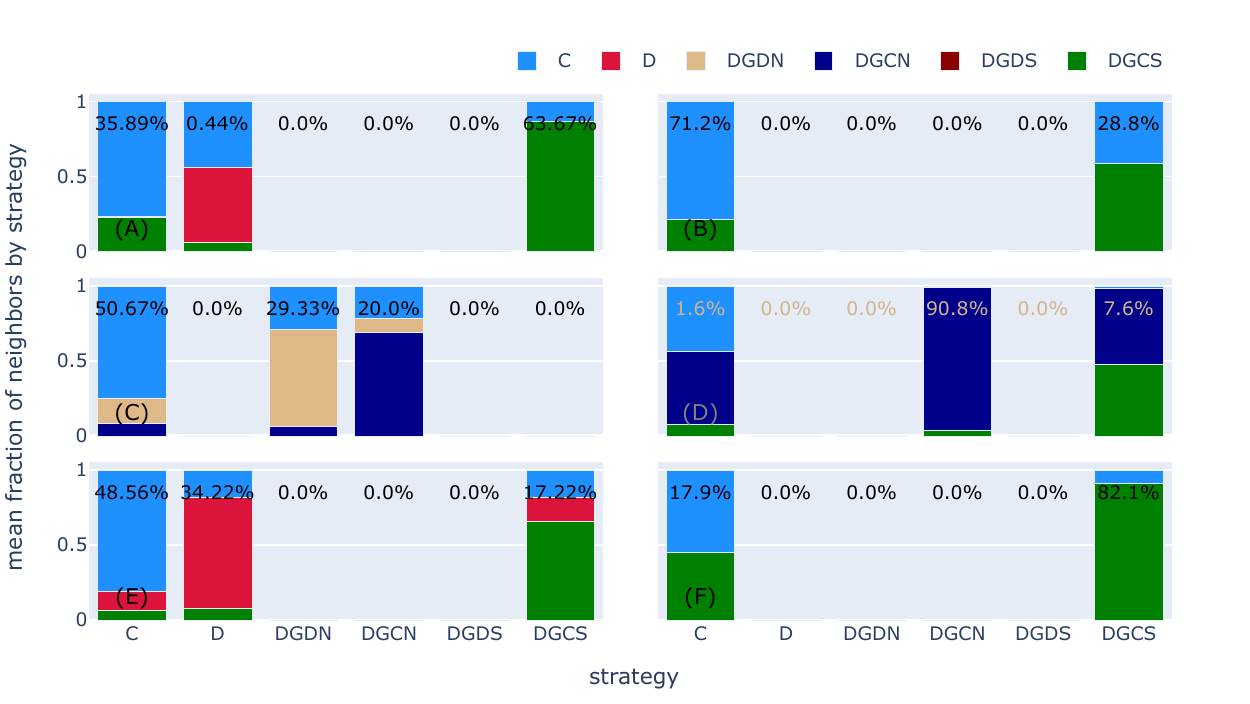}
    \caption{\textbf{Structured populations foster clustering in mixed strategy outcomes}. The stacked bars represent the mean fraction of strategists in the neighbourhood for each focal strategist. The percentage shown on the bar represents the total fraction of those players in the population. Left column reports results for square lattice ($N = 900$) and right one for scale-free networks ($N = 1000$). Typical runs selected to show mixed strategy outcomes if available (more replicates and different parameter values in SI).  Parameters: $\Omega = 10$; $b = 2$, $\gamma = 4$, $\gamma_s = 0$ (A and B); $b = 4$, $\gamma = 1$, $\gamma_s = 1$ (C and D); $b = 4$, $\gamma = 7$, $\gamma_s = 0$ (E and F).} 
    \label{fig:clustering}
\end{figure*}

Importantly, we see a shift in the cyclic dynamics previously encountered in well-mixed populations. This property can be clarified by observing the clustering behaviours typical of structured populations, even in the case of homogeneous graphs (see Figure \ref{fig:clustering}, left column). Typically, we see that unemotional cooperators (C) are better protected against unemotional defectors (D) when spatiality allows for network reciprocity, especially when evolutionary dynamics lead to mixed strategy outcomes (no one strategy fully dominates the others). Through such clusters, emotionally adaptive strategists (DGCN and DGCS) can often survive in the face of D players. Moreover, this can allow for the co-existence of guilt-prone individuals in communities of other like-minded strategists and  C players, especially if the cost of being social ($\gamma_s$) is low enough (e.g., $\gamma_s = 0$ and $\gamma_s = 1$, as highlighted in Figure \ref{fig:clustering}).

We now consider a more complex network structure, the scale-free (SF) network, heterogeneous and highly diverse in the number and distribution of connections. Previous works studying the evolution of cooperation on different networks showed that SF properties can markedly promote cooperation in one-shot social dilemmas, as heterogeneity in the network structure allows cooperators to form clusters around highly connected nodes (hubs) \citep{santos:2008:nature,santos:2005:prl,Szabo2007}. 
Our aim is to study whether this property would also allow pro-social behaviours to evolve; that is, strategies which would not have had a chance to do so previously. To this end, we investigate whether non-social guilt strategies can emerge, leading to even higher levels of (less-costly) cooperation overall. 

\begin{figure*}[h]
    \centering
    \includegraphics[width=\linewidth]{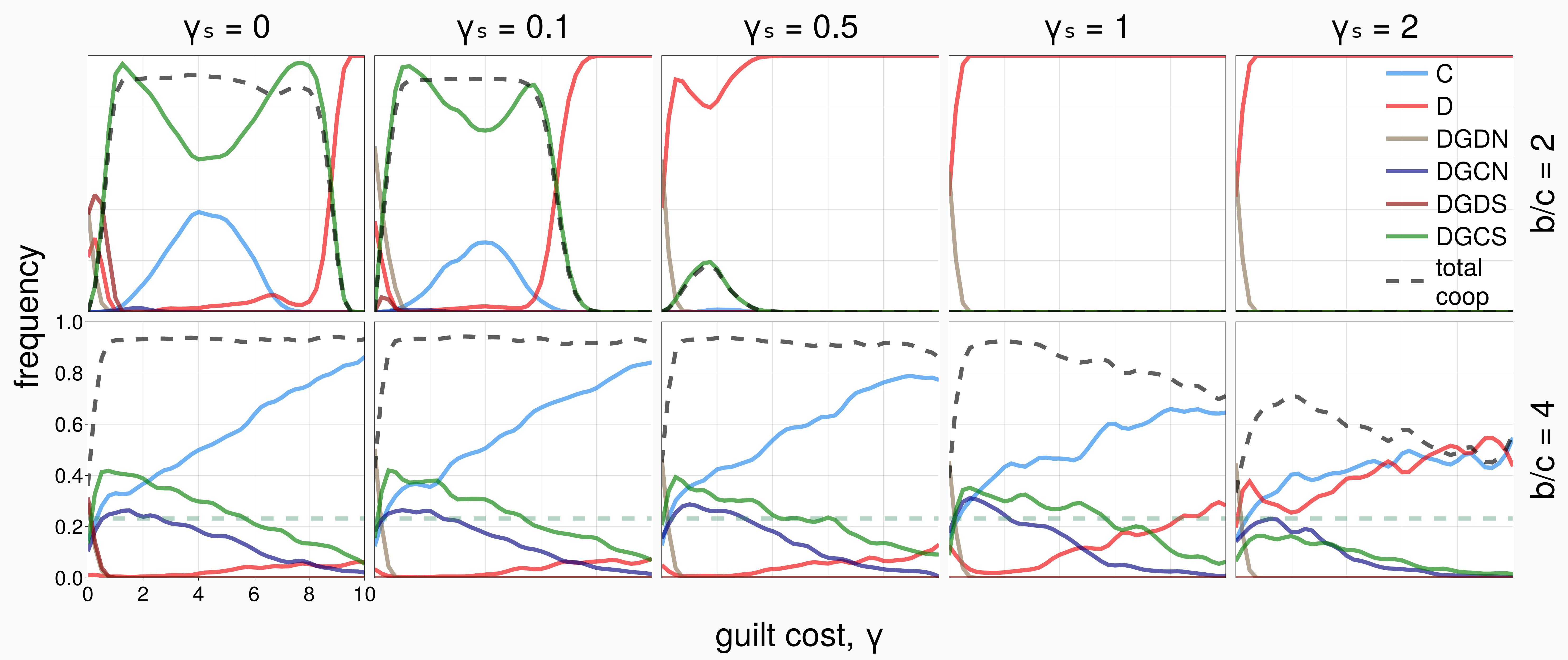}
    \caption{ Strategies' frequency and the total cooperation level as a function of the guilt cost, $\gamma$ (scale-free, $N = 1000$, $\Omega = 10$). When shown, the dashed green line marks the baseline level of cooperation achieved solely through network reciprocity.}
    \label{fig:gamma_sf}
\end{figure*}

We observe similar outcomes to SL  when $b = 2$, with a slight decrease of cooperation when $\gamma_s = 0.5$ (see Figure \ref{fig:gamma_sf}). When $b = 4$, we find higher levels of cooperation in SF than in SL, across a wide range of guilt and social costs. This improvement can be attributed to the success of non-social guilt, which becomes rather abundant across the entire parameter space. This is a remarkable observation, whereby the easily exploitable non-social individuals (which are nevertheless also desirably cost efficient) can evolve and co-exist with other strategies in an evolving population/MAS of self-interested agents. 

To further explain this finding and confirm our intuitions, we show the clustering behaviours typical of scale-free populations in Figure \ref{fig:clustering}, right column. Given a low social cost $\gamma_s$, social guilt can thrive even in cases when the cost of guilt $\gamma$ is very large (see Figure \ref{fig:clustering} panels B and F). Communities of emotionally adaptive individuals co-evolve and co-exist, surviving in the face of the predictions of evolutionary dynamics in homogeneous populations. That is, emotionally sacrificial strategies are empowered through heterogeneous environments, even in an incipient form that does not require costly monitoring of the surrounding contexts.

%Non-social guilt emerges in scale free network (through clustering with emotional strategies DGC1 and DGC2?), but not in lattice. 
%So, social inequality enables non-social guilt to evolve and co-exist with other strategies; 

% for each type, record their frequency of connections with all other  strategies in the equi state. 

\section{Discussion}
Based on  psychological and evolutionary accounts of guilt and social emotions, the present paper studies an evolutionary game theoretical model with social and non-social guilt-prone strategies in co-presence, in the context of differently structured populations (or distributed MAS). 
The paper considered several important  population  structures, from homogeneous ones, in the forms of well-mixed and square lattices, to heterogeneous, scale-free networks, showing that the evolutionary outcomes of social and non-social guilt strategies are highly dependent on the underlying population structure.   
We showed, in the context of the Iterated Prisoner's Dilemma, that only social guilt can evolve in the well-mixed population context, which is in line with previous findings in the literature \citep{Luis2017AAMAS} (see Supporting Information for additional analyses where social and non-social guilt strategies are considered separately). 
Spatial structures, even homogeneous ones (e.g. square lattices), allow guilt-prone strategies and cooperation to prevail for a much wider range of the guilt and social costs (compared to the well-mixed setting). Interestingly, heterogeneous networks (i.e. scale-free), and to a lesser extent square lattices, allow non-social guilt to evolve through the formation of clusters with other emotional  agents to defend against exploitation. 
This finding is remarkable, as it showed that costly guilt-prone strategies can prevail in spatial  environments, even in an incipient form which does not require expensive monitoring of the context behind others' actions. This is especially true when the underlying networks mirror realistic, heterogeneous structures \citep{barabasi2014linked}.

%\section{Related Work}
The problems of explaining the evolution and emergence of collective behaviours, such as cooperation, coordination and AI safety  in dynamical populations or systems of self-interested agents, have been actively studied across disciplines, from Evolutionary Biology, Physics, Economics, to AI and Multi-Agent Systems \citep{HanBook2013,tuyls2007evolutionary,cimpeanu2021cost,HanJaamas2016,merhej2022cooperation,xu2019cooperation,perc2017statistical,phelps2010evolutionary,han2019modelling,savarimuthu2011norm,santos2020picky,domingos2017reactive,KrellnerArtLife2022,Han2017AAAI,ogbo2022evolution}. Several mechanisms have been proposed to explain the dilemmas of cooperation, including kin selection, direct and indirect reciprocity, incentives or networked structures; see surveys in \citep{nowak:2006bo,sigmund:2010bo,perc2017statistical,Xia2023}. In contrast, there is a significant lack of studies looking at the role of cognitive and emotional mechanisms in behavioural evolution \citep{HanAICOM2022,dafoe2021cooperative,andras2018trusting,HanetalTrust2021,correia2025evolution}. Acknowledging the pivotal role of emotions on human decision-making \citep{marsella2014computationally,Turrini2010}, it is essential to incorporate these complex mechanisms for a more holistic portrayal of the evolution of cooperation. Our work strives to bridge this gap, providing key insights into the design and engineering of self-organized and distributed Multi-Agent Systems (MAS), especially in the context of a hybrid human-AI setting, such as cooperative AI \citep{paiva2018engineering,dafoe2021cooperative,andras2018trusting,HanetalTrust2021,zimmaro2024emergence,si2025cooperative,hammond2025multiagentrisksadvancedai}.

Most relevant to our work is the EGT model proposed in \citep{Luis2017AAMAS}, showing that cooperation does not emerge when agents only alleviate their own guilt (i.e. non-social guilt), without considering their co-players’ own attitudes about alleviation of guilt as well. That is the case where guilt-prone agents are easily dominated by agents who do not express guilt or who have no motivation to alleviate their own guilt. Hence, only when the tendency to alleviate guilt is mutual (i.e. social guilt),   can cooperation thrive.
This previous work did not consider that choosing to be social  might require  a cost (compared to being non-social), and thus the latter might have an evolutionary advantage against the former. Indeed, our (risk-dominance) analysis  shows that in a direct competition, a non-social guilt strategy is risk-dominant or advantageous against a social one. Because  this prior work did not consider both guilt-prone strategies in co-presence within a population, it was not possible to address how this social cost might affect the evolutionary outcomes. The present work considers an extended model where all these strategies are in co-presence together with other non-emotional strategies  in a population, so as to address these issues.   
Moreover, the prior work \citep{Luis2017AAMAS} only focused on the well-mixed population setting, therefore failing to assess how the structure of the underlying network of contacts among the agents in the population affects the evolutionary outcome and the design of cooperative societies. For example, our results show that a spatial structure, even if homogeneous like square lattices, allows guilt-prone strategies and cooperation to prevail for a much wider range of the guilt and social costs (compared to the  well-mixed setting). Heterogeneous (scale-free) networks, and to some extent square lattices, allow non-social guilt to evolve through clustering of guilt-prone individuals to avoid their exploiters.

Guilt has been considered implicitly in prior EGT models studying apology and forgiveness in social dilemma games \citep{ijcai2013TAH,martinez2015apology,oconnor,martinez2017agreement}. These works do not look at  guilt as part of  agents' strategies, but rather it plays an implicit role leading agents to make an apology after wrongdoings. In our present work, the modelling of guilt as a behavioural feature of a strategy enables exploration of new aspects related to feeling guilty, namely its social aspects and how it interacts with external factors, like the network's structure.

Our modelling work is inspired by a large number of works from psychological, sociological and philosophical literature. 
\citep{Ramsey2022} argue  that the evolutionary emergence of the emotion of guilt needs support on the evolution of empathy. From a multi-agent perspective, including mixed social-technological communities encompassing potentially autonomous artificial agents, and invoking the so-called “value alignment” problem (for a recent review cf. \citep{Gabriel2020}). In line with \citep{Luis2017AAMAS}, the outcomes from our analyses  help confirm that conflicts can be avoided when morally salient emotions, like guilt, help guide participants toward acceptable behaviours. In this context, systems involving possible future artificial moral agents may be designed to include guilt, to align agent-level behaviour with human expectations, thereby resulting in overall social benefits through improved cooperation.

Finally, there exists a large body  of computational modelling works of guilt in AI and MAS literature \citep{de2012bayesian,Fix2006,Savarimuthu2008,key:HanLpar2012,Turrini2010,criado2011open,o2016evolution,oconnor}.
Unlike our intended outcome, these studies are geared towards the formalization of guilt within Multi-Agent Systems (MAS), including virtual and cognitive agent systems. The purposes range from regulating social norms \citep{criado2011open} to improving agent decision-making and reasoning processes \citep{Turrini2010,marsella2014computationally}. Beyond that, our results  provide novel insights into the design and engineering of such  MAS systems; for instance, if agents are equipped with the capacity of guilt feeling even if it might lead to costly disadvantage, that can drive the system to an overall more cooperative outcome where they are willing to take reparative actions after wrongdoings.  
Additionally, our analysis provides insights on how such guilt-capable agents should be distributed to optimise cooperative outcomes, depending on the specific MAS network structure   \citep{marsella2014computationally,Savarimuthu2008,Turrini2010}.

To be evolutionarily viable, an advantageous guilt-prone agent-genotype must act in view of the capacity for its game partners to also express guilt, for a diversity of network structures. The lesson from these experiments is that self-punishment by suffering guilt, without considering whether partners are also similarly guilt-affected, does not result in guilt becoming a dominant advantageous feature of individuals in the population. On the contrary, when defecting partners do not express guilt when agents themselves do, then an agent should either not experience guilt or its guilt should be automatically alleviated, at no cost. Otherwise, guilt-prone agents would be exploited by the non-guilt prone free-riders with respect to guilt.

Within the Iterated Prisoner's Dilemma (IPD), agents assess each other's actions, deciding whether to defect or cooperate. In real-world scenarios, humans similarly take into account the thought processes that lead others to make these decisions. People, first, tend to trust others who cooperate without ever thinking about defecting over those who do consider defection  an option, and only later choose against trusting them. According to Kant, “In law a man is guilty when he violates the rights of others. In ethics he is guilty if he only thinks of doing so.” \citep{Hoffman2016}. 

Being attuned to the thought processes or behavioral indications of individuals contemplating cheating or deception entails an added ability to recognize intentions. In accordance with Kant's insights, \citet{Luis2017AAMAS} affirm that intention recognition plays a crucial role in regulating social interactions, even in cases where a given intention is not explicitly acted upon.
However, common sense stresses that feeling guilt for harm done to others only makes sense if one perceives those others did not intend harm and will feel guilty for harm done as well. 
Where recognizing the intention of another is not considered, then feeling guilty about defections without regard to what others conceivably feel about their defections is self-defeating.
 %LINK TO FUTURE WORK ON INTENTION RECOGNITION EXTENSION AND CITE OUR PREVIOUS IR WORK.

In essence, the present research has provided a robust, game-theoretical-based account of how the interplay between social costs and underlying network structures in a population or distributed Multi-Agent System (MAS) enables the co-evolution and coexistence of different types of social and non-social emotions. As a desired result, this strengthens cooperation, though their beholders will incur a significant emotional cost to themselves to achieve this. 

\backmatter

\section*{Ethics}

This work did not require ethical approval from a human subject or animal welfare committee.

\section*{Funding}

T.C. is supported by the UKRI CRCRM (MR/Z505833/1).
T.A.H. is supported by EPSRC (grant EP/Y00857X/1).

 \section*{Authors' Contribution}

 \textbf{T.C.:} Conceptualization (equal); data curation (lead); formal analysis (equal); investigation (lead); Methodology (equal); Software (lead); Validation (equal); Visualization (lead); writing --  original draft, review and editing (equal). \textbf{L.M.P.:} Conceptualization (equal);
 Validation (equal); writing -- original draft, review and editing (equal). \textbf{T.A.H.:} Conceptualization (equal); formal analysis (equal); investigation (supporting); Methodology (equal); Validation (equal); Visualization (supporting); writing -- original draft, review and editing (equal).

\section*{Data accessibility}
The data, code and supplmentary material that support the findings of this study are available at \url{http://datadryad.org/stash/share/3r-Ef414DYUhD6NC74_YqDDfQbq4SzuY0qWs9vdsRtM}. 

%\begin{appendices}

\bibliography{sample}% Produces the bibliography via BibTeX.

\begin{thebibliography}{91}
\providecommand{\natexlab}[1]{#1}
\providecommand{\url}[1]{{#1}}
\providecommand{\urlprefix}{URL }
\providecommand{\doi}[1]{\url{https://doi.org/#1}}
\providecommand{\eprint}[2][]{\url{#2}}
 \bibcommenthead

\bibitem[{Andras et~al(2018)Andras, Esterle, Guckert, Han, Lewis, Milanovic,
  Payne, Perret, Pitt, Powers et~al}]{andras2018trusting}
Andras P, Esterle L, Guckert M, et~al (2018) Trusting intelligent machines:
  Deepening trust within socio-technical systems. IEEE Technology and Society
  Magazine 37(4):76--83

\bibitem[{Awad et~al(2018)Awad, Dsouza, Kim, Schulz, Henrich, Shariff,
  Bonnefon, and Rahwan}]{Awad2018}
Awad E, Dsouza S, Kim R, et~al (2018) The moral machine experiment. Nature
  563:59--64. \doi{10.1038/s41586-018-0637-6},
  \urlprefix\url{https://doi.org/10.1038/s41586-018-0637-6}

\bibitem[{Barabasi(2014)}]{barabasi2014linked}
Barabasi AL (2014) Linked-how Everything is Connected to Everything Else and
  what it Means F. Perseus Books Group

\bibitem[{Barab{\'{a}}si(2016)}]{Barabasi2016}
Barab{\'{a}}si AL (2016) {Network Science}. Cambridge University Press

\bibitem[{Barab{\'a}si and Albert(1999)}]{barabasi1999emergence}
Barab{\'a}si AL, Albert R (1999) Emergence of scaling in random networks.
  science 286(5439):509--512

\bibitem[{Bastin et~al(2016)Bastin, Harrison, Davey, Moll, and
  Whittle}]{Bastin2016}
Bastin C, Harrison BJ, Davey CG, et~al (2016) {Feelings of shame, embarrassment
  and guilt and their neural correlates: A systematic review}. Neuroscience
  {\&} Biobehavioral Reviews 71:455--471.
  \doi{https://doi.org/10.1016/j.neubiorev.2016.09.019},
  \urlprefix\url{https://www.sciencedirect.com/science/article/pii/S0149763415302876}

\bibitem[{Billingham and Parr(2020)}]{Billingham2020}
Billingham P, Parr T (2020) {Online Public Shaming: Virtues and Vices}. Journal
  of Social Philosophy 51(3):371--390.
  \doi{https://doi.org/10.1111/josp.12308},
  \urlprefix\url{https://doi.org/10.1111/josp.12308}

\bibitem[{Burnett et~al(2009)Burnett, Bird, Moll, Frith, and
  Blakemore}]{burnett2009development}
Burnett S, Bird G, Moll J, et~al (2009) Development during adolescence of the
  neural processing of social emotion. Journal of cognitive neuroscience
  21(9):1736--1750

\bibitem[{Cimpeanu et~al(2021)Cimpeanu, Perret, and Han}]{cimpeanu2021cost}
Cimpeanu T, Perret C, Han TA (2021) {Cost-efficient interventions for promoting
  fairness in the ultimatum game}. Knowledge-Based Systems 233:107545

\bibitem[{Cimpeanu et~al(2023)Cimpeanu, Di~Stefano, Perret, and
  Han}]{cimpeanu2023social}
Cimpeanu T, Di~Stefano A, Perret C, et~al (2023) Social diversity reduces the
  complexity and cost of fostering fairness. Chaos, Solitons \& Fractals
  167:113051

\bibitem[{Conradi(2010)}]{Conradi2010}
Conradi PJ (2010) {Laughing at Something Tragic: Murdoch as Anti-Moralist BT -
  Iris Murdoch and Morality}. Palgrave Macmillan UK, London, p 56--69,
  \doi{10.1057/9780230277229_5},
  \urlprefix\url{https://doi.org/10.1057/9780230277229{\_}5}

\bibitem[{Coombs(1973)}]{coombs1973reparameterization}
Coombs CH (1973) A reparameterization of the prisoner's dilemma game.
  Behavioral Science 18(6):424--428

\bibitem[{Criado et~al(2011)Criado, Argente, and Botti}]{criado2011open}
Criado N, Argente E, Botti V (2011) Open issues for normative multi-agent
  systems. AI Communications 24(3):233--264

\bibitem[{Dafoe et~al(2021)Dafoe, Bachrach, Hadfield, Horvitz, Larson, Graepel
  et~al}]{dafoe2021cooperative}
Dafoe A, Bachrach Y, Hadfield G, et~al (2021) Cooperative ai: machines must
  learn to find common ground. Nature 593(7857):33--36

\bibitem[{De~Hooge et~al(2010)De~Hooge, Zeelenberg, and
  Breugelmans}]{de2010restore}
De~Hooge IE, Zeelenberg M, Breugelmans SM (2010) Restore and protect
  motivations following shame. Cognition and Emotion 24(1):111--127

\bibitem[{De~Melo et~al(2012)De~Melo, Carnevale, Read, Antos, and
  Gratch}]{de2012bayesian}
De~Melo CM, Carnevale P, Read S, et~al (2012) Bayesian model of the social
  effects of emotion in decision-making in multiagent systems. In: AAMAS'2012,
  pp 55--62

\bibitem[{Domingos et~al(2017)Domingos, Burguillo, and
  Lenaerts}]{domingos2017reactive}
Domingos EF, Burguillo JC, Lenaerts T (2017) Reactive versus anticipative
  decision making in a novel gift-giving game. In: Thirty-First AAAI Conference
  on Artificial Intelligence, pp 4399--4405

\bibitem[{Fessler and Haley(2003)}]{Daniel2003}
Fessler D, Haley KJ (2003) The strategy of affect: Emotions in human
  cooperation 12. The Genetic and Cultural Evolution of Cooperation, P
  Hammerstein, ed pp 7--36

\bibitem[{Fix et~al(2006)Fix, von Scheve, and Moldt}]{Fix2006}
Fix J, von Scheve C, Moldt D (2006) Emotion-based norm enforcement and
  maintenance in multi-agent systems: Foundations and petri net modeling. In:
  AAMAS '06. ACM, pp 105--107

\bibitem[{Flores and Han(2024)}]{flores2024evolution}
Flores LS, Han TA (2024) Evolution of commitment in the spatial public goods
  game through institutional incentives. Applied Mathematics and Computation
  473:128646

\bibitem[{Correia~da Fonseca et~al(2025)Correia~da Fonseca, de~Melo, Terada,
  Gratch, Paiva, and Santos}]{correia2025evolution}
Correia~da Fonseca H, de~Melo CM, Terada K, et~al (2025) Evolution of indirect
  reciprocity under emotion expression. Scientific Reports 15(1):9151

\bibitem[{Frank(1998)}]{frank:1998bv}
Frank SA (1998) Foundations of social evolution. Princeton Univ. Press,
  Princeton

\bibitem[{Gabriel(2020)}]{Gabriel2020}
Gabriel I (2020) {Artificial Intelligence, Values, and Alignment}. Minds and
  Machines 30(3):411--437. \doi{10.1007/s11023-020-09539-2},
  \urlprefix\url{https://doi.org/10.1007/s11023-020-09539-2}

\bibitem[{Gaudou et~al(2014)Gaudou, Lorini, and Mayor}]{gaudou2014moral}
Gaudou B, Lorini E, Mayor E (2014) Moral guilt: An agent-based model analysis.
  In: Advances in social simulation. Springer, p 95--106

\bibitem[{Guo et~al(2023)Guo, Song, Perc, Li, and Wang}]{guo2023third}
Guo H, Song Z, Perc M, et~al (2023) Third-party intervention of cooperation in
  multilayer networks. IEEE Transactions on Systems, Man, and Cybernetics:
  Systems

\bibitem[{Hammond et~al(2025)Hammond, Chan, Clifton, Hoelscher-Obermaier, Khan,
  McLean, Smith, Barfuss, Foerster, Gavenčiak, Han, Hughes, Kovařík,
  Kulveit, Leibo, Oesterheld, de~Witt, Shah, Wellman, Bova, Cimpeanu, Ezell,
  Feuillade-Montixi, Franklin, Kran, Krawczuk, Lamparth, Lauffer, Meinke,
  Motwani, Reuel, Conitzer, Dennis, Gabriel, Gleave, Hadfield, Haghtalab,
  Kasirzadeh, Krier, Larson, Lehman, Parkes, Piliouras, and
  Rahwan}]{hammond2025multiagentrisksadvancedai}
Hammond L, Chan A, Clifton J, et~al (2025) Multi-agent risks from advanced ai.
  \urlprefix\url{https://arxiv.org/abs/2502.14143}, \eprint{2502.14143}

\bibitem[{Han(2013)}]{HanBook2013}
Han TA (2013) Intention Recognition, Commitments and Their Roles in the
  Evolution of Cooperation: From Artificial Intelligence Techniques to
  Evolutionary Game Theory Models, vol~9. Springer SAPERE series

\bibitem[{Han(2022)}]{HanAICOM2022}
Han TA (2022) Emergent behaviours in multi-agent systems with evolutionary game
  theory. AI Communications 35(4):327 – 337

\bibitem[{Han et~al(2012)Han, Saptawijaya, and Pereira}]{key:HanLpar2012}
Han TA, Saptawijaya A, Pereira LM (2012) Moral reasoning under uncertainty. In:
  Proceedings of the 18th International Conference on Logic for Programming,
  Artificial Intelligence and Reasoning (LPAR-18). Springer LNAI 7180, pp
  212--227

\bibitem[{Han et~al(2013)Han, Pereira, Santos, and Lenaerts}]{ijcai2013TAH}
Han TA, Pereira LM, Santos FC, et~al (2013) {Why Is It So Hard to Say Sorry:
  The Evolution of Apology with Commitments in the Iterated Prisoner's
  Dilemma}. In: IJCAI'2013. AAAI Press, pp 177--183

\bibitem[{Han et~al(2017{\natexlab{a}})Han, Pereira, and
  Lenaerts}]{HanJaamas2016}
Han TA, Pereira LM, Lenaerts T (2017{\natexlab{a}}) Evolution of commitment and
  level of participation in public goods games. Autonomous Agents and
  Multi-Agent Systems pp 1--23

\bibitem[{Han et~al(2017{\natexlab{b}})Han, Pereira, Martinez-Vaquero, and
  Lenaerts}]{Han2017AAAI}
Han TA, Pereira LM, Martinez-Vaquero LA, et~al (2017{\natexlab{b}}) Centralized
  vs. personalized commitments and their influence on cooperation in group
  interactions. In: AAAI, pp 2999--3005

\bibitem[{Han et~al(2020)Han, Pereira, Santos, and Lenaerts}]{han2019modelling}
Han TA, Pereira LM, Santos FC, et~al (2020) {To Regulate or Not: A Social
  Dynamics Analysis of an Idealised AI Race}. Journal of Artificial
  Intelligence Research 69:881--921

\bibitem[{Han et~al(2021)Han, Perret, and Powers}]{HanetalTrust2021}
Han TA, Perret C, Powers ST (2021) When to (or not to) trust intelligent
  machines: Insights from an evolutionary game theory analysis of trust in
  repeated games. Cognitive Systems Research 68:111--124

\bibitem[{Hareli and Parkinson(2008)}]{hareli2008s}
Hareli S, Parkinson B (2008) What's social about social emotions? Journal for
  the Theory of Social Behaviour 38(2):131--156

\bibitem[{Hofbauer and Sigmund(1998)}]{key:Hofbauer1998}
Hofbauer J, Sigmund K (1998) Evolutionary Games and Population Dynamics.
  Cambridge University Press

\bibitem[{Hoffman et~al(2016)Hoffman, Yoeli, and Navarrete}]{Hoffman2016}
Hoffman M, Yoeli E, Navarrete CD (2016) {Game Theory and Morality BT - The
  Evolution of Morality}. Springer International Publishing, Cham, p 289--316,
  \doi{10.1007/978-3-319-19671-8_14},
  \urlprefix\url{https://doi.org/10.1007/978-3-319-19671-8{\_}14}

\bibitem[{Imhof et~al(2005)Imhof, Fudenberg, and Nowak}]{key:imhof2005}
Imhof LA, Fudenberg D, Nowak MA (2005) Evolutionary cycles of cooperation and
  defection. Proc Natl Acad Sci USA 102:10797--10800

\bibitem[{Joyce(2007)}]{Joyce2006}
Joyce R (2007) The evolution of morality. MIT press

\bibitem[{Joyce(2008)}]{Joyce2008}
Joyce R (2008) 3.2 aversions, sentiments, moral judgments, and taboos. Moral
  Psychology: The Evolution of Morality: Adaptations and Innateness 1:195

\bibitem[{Keng and Tan(2017)}]{Keng2017}
Keng SL, Tan JX (2017) {Effects of brief mindful breathing and loving-kindness
  meditation on shame and social problem solving abilities among individuals
  with high borderline personality traits}. Behaviour Research and Therapy
  97:43--51. \doi{https://doi.org/10.1016/j.brat.2017.07.004},
  \urlprefix\url{https://www.sciencedirect.com/science/article/pii/S0005796717301389}

\bibitem[{Ketelaar and Tung~Au(2003)}]{KetelaanAu2003}
Ketelaar T, Tung~Au W (2003) The effects of feelings of guilt on the behaviour
  of uncooperative individuals in repeated social bargaining games: An
  affect-as-information interpretation of the role of emotion in social
  interaction. Cognition and emotion 17(3):429--453

\bibitem[{Kowalczuk and Czubenko(2016)}]{kowalczuk2016computational}
Kowalczuk Z, Czubenko M (2016) Computational approaches to modeling artificial
  emotion--an overview of the proposed solutions. Frontiers in Robotics and AI
  3:21

\bibitem[{Krellner and Han(2022)}]{KrellnerArtLife2022}
Krellner M, Han TA (2022) {Pleasing Enhances Indirect Reciprocity-Based
  Cooperation Under Private Assessment}. Artificial Life 27(3–4):246--276

\bibitem[{Köbis et~al(2021)Köbis, Bonnefon, and Rahwan}]{kobis2021}
Köbis N, Bonnefon JF, Rahwan I (2021) Bad machines corrupt good morals. Nature
  Human Behaviour 5:679--685. \doi{10.1038/s41562-021-01128-2},
  \urlprefix\url{https://doi.org/10.1038/s41562-021-01128-2}

\bibitem[{Man and Damasio(2019)}]{man2019homeostasis}
Man K, Damasio A (2019) Homeostasis and soft robotics in the design of feeling
  machines. Nature Machine Intelligence 1(10):446--452

\bibitem[{Marsella and Gratch(2014)}]{marsella2014computationally}
Marsella S, Gratch J (2014) Computationally modeling human emotion.
  Communications of the ACM 57(12):56--67

\bibitem[{Martinez-Vaquero et~al(2015)Martinez-Vaquero, Han, Pereira, and
  Lenaerts}]{martinez2015apology}
Martinez-Vaquero LA, Han TA, Pereira LM, et~al (2015) Apology and forgiveness
  evolve to resolve failures in cooperative agreements. Scientific reports
  5(10639)

\bibitem[{Martinez-Vaquero et~al(2017)Martinez-Vaquero, Han, Pereira, and
  Lenaerts}]{martinez2017agreement}
Martinez-Vaquero LA, Han TA, Pereira LM, et~al (2017) When agreement-accepting
  free-riders are a necessary evil for the evolution of cooperation. Scientific
  reports 7(1):2478

\bibitem[{Mayer and Vanderheiden(2021)}]{Mayer2021}
Mayer CH, Vanderheiden E (2021) {Naming and Shaming in Cyberspace: Forms,
  Effects and Counterstrategies BT - Shame 4.0: Investigating an Emotion in
  Digital Worlds and the Fourth Industrial Revolution}. Springer International
  Publishing, Cham, p 389--412, \doi{10.1007/978-3-030-59527-2_18},
  \urlprefix\url{https://doi.org/10.1007/978-3-030-59527-2{\_}18}

\bibitem[{Merhej et~al(2022)Merhej, Santos, Melo, and
  Santos}]{merhej2022cooperation}
Merhej R, Santos FP, Melo FS, et~al (2022) Cooperation and learning dynamics
  under wealth inequality and diversity in individual risk. J Artif Int Res 74.
  \doi{10.1613/jair.1.13519},
  \urlprefix\url{https://doi.org/10.1613/jair.1.13519}

\bibitem[{Nesse(2019)}]{Nesse2019}
Nesse RM (2019) {Good Reasons for Bad Feelings: Insights from the Frontier of
  Evolutionary Psychiatry}. Allen Lane

\bibitem[{Nowak(2006)}]{nowak:2006bo}
Nowak MA (2006) Evolutionary Dynamics. Harvard University Press, Cambridge, MA

\bibitem[{Nowak et~al(2004)Nowak, Sasaki, Taylor, and
  Fudenberg}]{key:novaknature2004}
Nowak MA, Sasaki A, Taylor C, et~al (2004) Emergence of cooperation and
  evolutionary stability in finite populations. Nature 428:646--650

\bibitem[{Ogbo et~al(2022)Ogbo, Elragig, and Han}]{ogbo2022evolution}
Ogbo NB, Elragig A, Han TA (2022) Evolution of coordination in pairwise and
  multi-player interactions via prior commitments. Adaptive Behavior
  30(3):257--277

\bibitem[{Ohtsuki et~al(2007)Ohtsuki, Nowak, and Pacheco}]{ohtsuki2007breaking}
Ohtsuki H, Nowak MA, Pacheco JM (2007) {Breaking the symmetry between
  interaction and replacement in evolutionary dynamics on graphs}. Physical
  review letters 98(10):108106

\bibitem[{O’Connor(2016)}]{o2016evolution}
O’Connor C (2016) The evolution of guilt: a model-based approach. Philosophy
  of Science 83(5):897--908

\bibitem[{Paiva et~al(2018)Paiva, Santos, and Santos}]{paiva2018engineering}
Paiva A, Santos FP, Santos FC (2018) Engineering pro-sociality with autonomous
  agents. In: Thirty-second AAAI conference on artificial intelligence

\bibitem[{Perc et~al(2017)Perc, Jordan, Rand, Wang, Boccaletti, and
  Szolnoki}]{perc2017statistical}
Perc M, Jordan JJ, Rand DG, et~al (2017) Statistical physics of human
  cooperation. Phys Rep 687:1--51

\bibitem[{Pereira and Santos(2019)}]{Pereira2019}
Pereira LM, Santos FC (2019) Counterfactual thinking in cooperation dynamics.
  In: Model-Based Reasoning in Science and Technology: Inferential Models for
  Logic, Language, Cognition and Computation, Springer, pp 69--82

\bibitem[{Pereira et~al(2016)Pereira, Saptawijaya
  et~al}]{pereira2016programming}
Pereira LM, Saptawijaya A, et~al (2016) Programming machine ethics, vol~26.
  Springer

\bibitem[{Pereira et~al(2017)Pereira, Lenaerts, Martinez-Vaquero, and
  Han}]{Luis2017AAMAS}
Pereira LM, Lenaerts T, Martinez-Vaquero LA, et~al (2017) Social manifestation
  of guilt leads to stable cooperation in multi-agent systems. In: AAMAS, pp
  1422--1430

\bibitem[{Pereira et~al(2021)Pereira, Han, and Lopes}]{pereira2021employing}
Pereira LM, Han TA, Lopes AB (2021) Employing ai to better understand our
  morals. Entropy 24(1):10

\bibitem[{Pereira et~al(2024)Pereira, Santos, and Lopes}]{Pereira2024}
Pereira LM, Santos FC, Lopes AB (2024) AI Modelling of Counterfactual Thinking
  for Judicial Reasoning and Governance of Law, Springer International
  Publishing, Cham, pp 263--279. \doi{10.1007/978-3-031-41264-6_14},
  \urlprefix\url{https://doi.org/10.1007/978-3-031-41264-6_14}

\bibitem[{Phelps et~al(2010)Phelps, McBurney, and
  Parsons}]{phelps2010evolutionary}
Phelps S, McBurney P, Parsons S (2010) Evolutionary mechanism design: a review.
  Autonomous Agents and Multi-Agent Systems 21(2):237--264

\bibitem[{Prinz(2008)}]{Prinz2008}
Prinz J (2008) Is morality innate. Moral psychology 1:367--406

\bibitem[{Prinz and Nichols(2010)}]{Prinz2010}
Prinz JJ, Nichols S (2010) 1. the role of emotions in moral cognition. The
  Moral Psychology Handbook p 111

\bibitem[{Ramsey and Deem(2022)}]{Ramsey2022}
Ramsey G, Deem MJ (2022) {Empathy and the Evolutionary Emergence of Guilt}.
  Philosophy of Science 89(3):434--453. \doi{DOI: 10.1017/psa.2021.36},
  \urlprefix\url{https://www.cambridge.org/core/article/empathy-and-the-evolutionary-emergence-of-guilt/861EED90948E1575494128853B5B6854}

\bibitem[{Rand et~al(2013)Rand, Tarnita, Ohtsuki, and Nowak}]{randUltimatum}
Rand DG, Tarnita CE, Ohtsuki H, et~al (2013) Evolution of fairness in the
  one-shot anonymous ultimatum game. Proc Natl Acad Sci USA 110:2581--2586

\bibitem[{Rosenstock and {O'Connor}(2016)}]{oconnor}
Rosenstock S, {O'Connor} C (2016) When it's good to feel bad: Evolutionary
  models of guilt and apology. Philosophy of Science 64(6):637--658

\bibitem[{Santos and Pacheco(2005)}]{santos:2005:prl}
Santos FC, Pacheco JM (2005) Scale-free networks provide a unifying framework
  for the emergence of cooperation. Phys Rev Lett 95:098104

\bibitem[{Santos et~al(2008)Santos, Santos, and Pacheco}]{santos:2008:nature}
Santos FC, Santos MD, Pacheco JM (2008) Social diversity promotes the emergence
  of cooperation in public goods games. Nature 454:214--216

\bibitem[{Santos et~al(2020)Santos, Mascarenhas, Santos, Correia, Gomes, and
  Paiva}]{santos2020picky}
Santos FP, Mascarenhas S, Santos FC, et~al (2020) Picky losers and carefree
  winners prevail in collective risk dilemmas with partner selection.
  Autonomous Agents and Multi-Agent Systems 34(2):1--29

\bibitem[{Savarimuthu and Cranefield(2011)}]{savarimuthu2011norm}
Savarimuthu BTR, Cranefield S (2011) Norm creation, spreading and emergence: A
  survey of simulation models of norms in multi-agent systems. Multiagent and
  Grid Systems 7(1):21--54

\bibitem[{Savarimuthu et~al(2008)Savarimuthu, Purvis, and
  Purvis}]{Savarimuthu2008}
Savarimuthu BTR, Purvis M, Purvis M (2008) Social norm emergence in virtual
  agent societies. In: AAMAS '08, pp 1521--1524

\bibitem[{Si et~al(2025)Si, He, Shen, and Tanimoto}]{si2025cooperative}
Si Z, He Z, Shen C, et~al (2025) Cooperative bots exhibit nuanced effects on
  cooperation across strategic frameworks. Journal of the Royal Society
  Interface 22(222):20240427

\bibitem[{Sigmund(2010{\natexlab{a}})}]{sigmund:2010bo}
Sigmund K (2010{\natexlab{a}}) The calculus of selfishness. Princeton Univ.
  Press

\bibitem[{Sigmund(2010{\natexlab{b}})}]{key:Sigmund_selfishnes}
Sigmund K (2010{\natexlab{b}}) The Calculus of Selfishness. Princeton
  University Press

\bibitem[{Szab{\'o} and F{\'a}th(2007)}]{Szabo2007}
Szab{\'o} G, F{\'a}th G (2007) Evolutionary games on graphs. Phys Rep
  97-216(4-6)

\bibitem[{Szab\'{o} et~al(2007)Szab\'{o}, Cz\'{a}r\'{a}n, and
  Szab\'{o}}]{szabo:2007jt}
Szab\'{o} P, Cz\'{a}r\'{a}n T, Szab\'{o} G (2007) Competing associations in
  bacterial warfare with two toxins. J theor Biol 248:736--744

\bibitem[{Tangney et~al(2013)Tangney, Stuewig, Malouf, and
  Youman}]{tangney201323}
Tangney JP, Stuewig J, Malouf ET, et~al (2013) 23 communicative functions of
  shame and guilt. Cooperation and its evolution p 485

\bibitem[{Tomasello(2016)}]{tomasello2016}
Tomasello M (2016) A natural history of human morality. Harvard University
  Press

\bibitem[{Traulsen et~al(2006)Traulsen, Nowak, and Pacheco}]{traulsen2006}
Traulsen A, Nowak MA, Pacheco JM (2006) Stochastic dynamics of invasion and
  fixation. Phys Rev E 74:11909

\bibitem[{Trivers(1971)}]{key:trivers1971}
Trivers RL (1971) The evolution of reciprocal altruism. Quaterly Review of
  Biology 46:35--57

\bibitem[{Turrini et~al(2010)Turrini, Meyer, and Castelfranchi}]{Turrini2010}
Turrini P, Meyer JJC, Castelfranchi C (2010) Coping with shame and sense of
  guilt: a dynamic logic account. Autonomous Agents and Multi-Agent Systems
  20(3):401--420. \doi{10.1007/s10458-009-9083-z},
  \urlprefix\url{http://dx.doi.org/10.1007/s10458-009-9083-z}

\bibitem[{Tuyls and Parsons(2007)}]{tuyls2007evolutionary}
Tuyls K, Parsons S (2007) What evolutionary game theory tells us about
  multiagent learning. Artificial Intelligence 171(7):406--416

\bibitem[{Xia et~al(2023)Xia, Wang, Perc, and Wang}]{Xia2023}
Xia C, Wang J, Perc M, et~al (2023) Reputation and reciprocity. Physics of Life
  Reviews 46:8--45. \doi{https://doi.org/10.1016/j.plrev.2023.05.002},
  \urlprefix\url{https://www.sciencedirect.com/science/article/pii/S1571064523000489}

\bibitem[{Xu et~al(2019)Xu, Garcia, and Handfield}]{xu2019cooperation}
Xu J, Garcia J, Handfield T (2019) Cooperation with bottom-up reputation
  dynamics. In: Proceedings of the 18th International Conference on Autonomous
  Agents and MultiAgent Systems, pp 269--276

\bibitem[{Zhou et~al(2025)Zhou, Zhu, Xia, and Chica}]{zhou2025}
Zhou C, Zhu Y, Xia C, et~al (2025) Evolutionary dynamics of trust in
  hierarchical populations with varying investment strategies. Journal of the
  Royal Society Interface 22. \doi{10.1098/RSIF.2024.0734},
  \urlprefix\url{https://royalsocietypublishing.org/doi/10.1098/rsif.2024.0734}

\bibitem[{Zimmaro et~al(2024)Zimmaro, Miranda, Fern{\'a}ndez, Moreno~L{\'o}pez,
  Reddel, Widler, Antonioni, and Han}]{zimmaro2024emergence}
Zimmaro F, Miranda M, Fern{\'a}ndez JMR, et~al (2024) Emergence of cooperation
  in the one-shot prisoner’s dilemma through discriminatory and samaritan
  ais. Journal of the Royal Society Interface 21(218):20240212

\bibitem[{Zisis et~al(2015)Zisis, Guida, Han, Kirchsteiger, and
  Lenaerts}]{zisisSciRep2015}
Zisis I, Guida SD, Han TA, et~al (2015) Generosity motivated by acceptance -
  evolutionary analysis of an anticipation games. Scientific reports 5(18076)

\end{thebibliography}

\end{document}

\title[The Evolutionary Advantage of Guilt]{The Evolutionary Advantage of Guilt: Supplementary Information}

\maketitle

\section{Supplementary material}
\renewcommand{\thefigure}{S\arabic{figure}}

This Appendix provides additional data results to support the robustness of the findings in the main paper. Here, we provide outcomes for the settings where social and non-social guilt strategies are considered separately, for well-mixed, square lattice and scale-free networks.

\begin{figure*}[h]
    \centering
    \includegraphics[width=\linewidth]{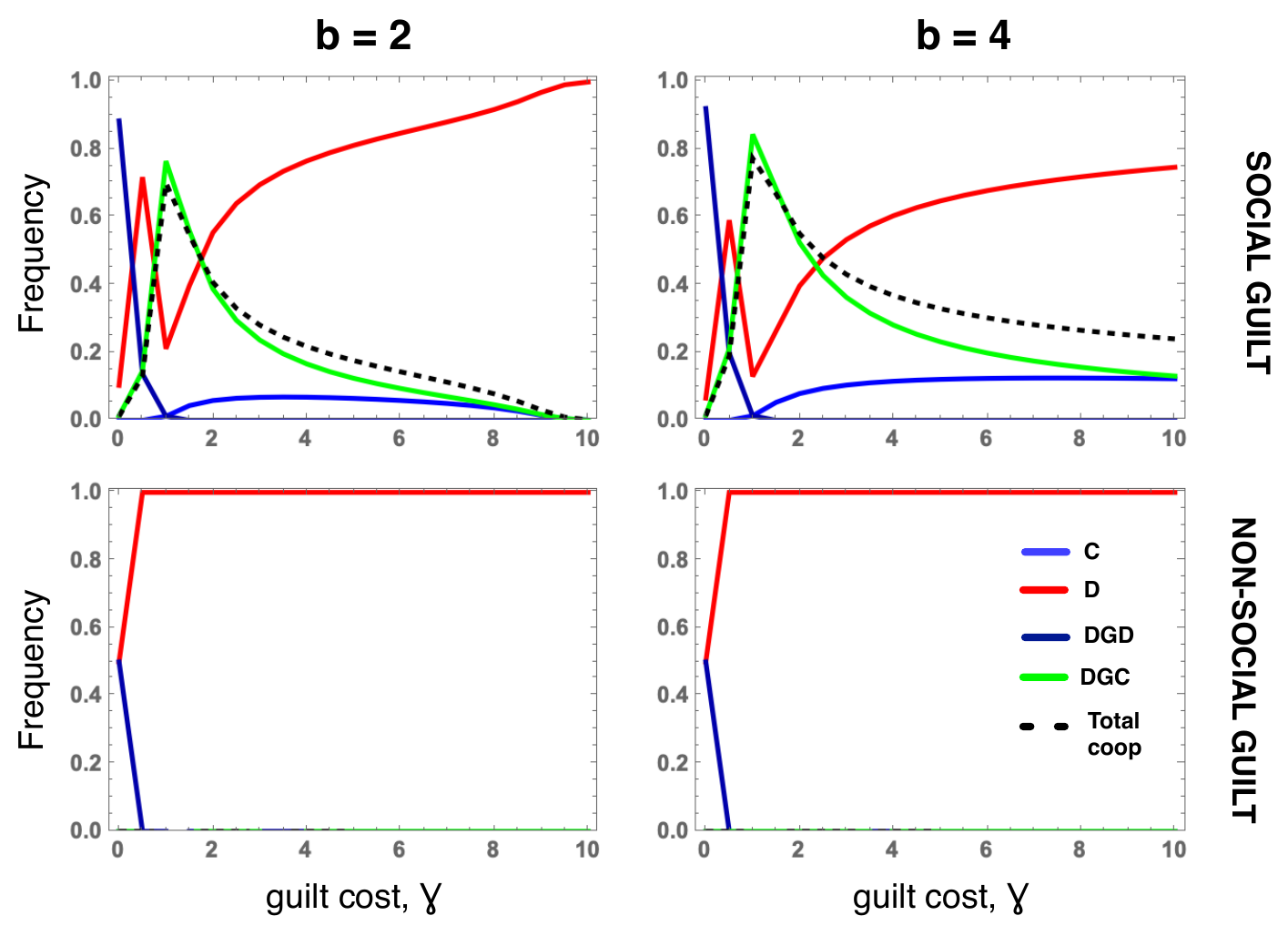}
    \caption{\textbf{Strategies' frequency and the total cooperation level as a function of the guilt cost, $\gamma$, for social vs non-social guilt}.     Other parameters: population size $N = 100$, $\Omega = 10$, $\beta = 1.0$. }
    \label{fig:dependgamma-appendix}
\end{figure*}

\begin{figure*}[h]
    \centering
    \includegraphics[width=\linewidth]{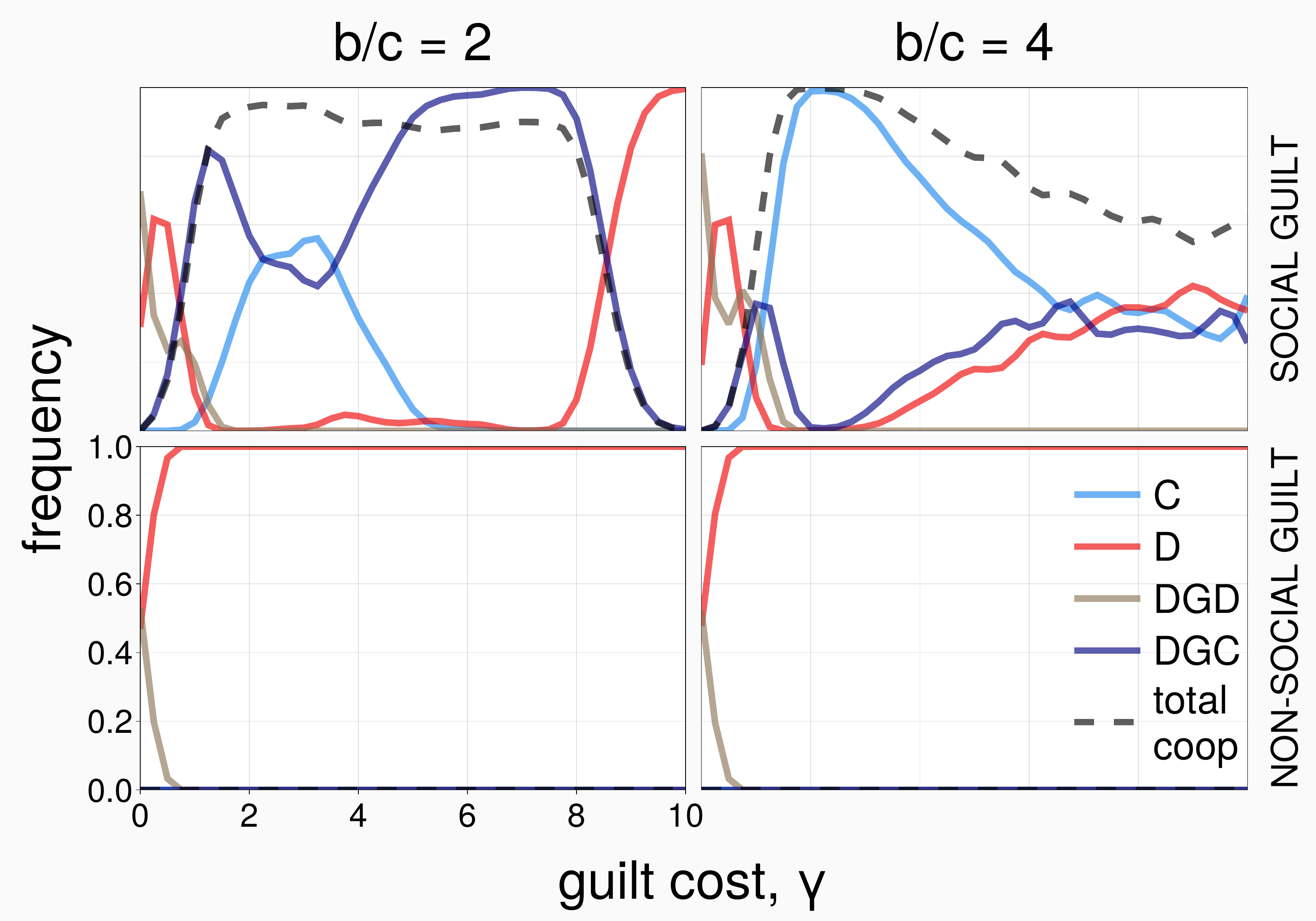}
    \caption{\textbf{Strategies' frequency and the total cooperation level as a function of the guilt cost, $\gamma$, for lattice networks}.     Other parameters: population size $N = 900$, $\Omega = 10$, $\beta = 1.0$. }
    \label{fig:gamma4strats_lattice}
\end{figure*}

\begin{figure*}[h]
    \centering
    \includegraphics[width=\linewidth]{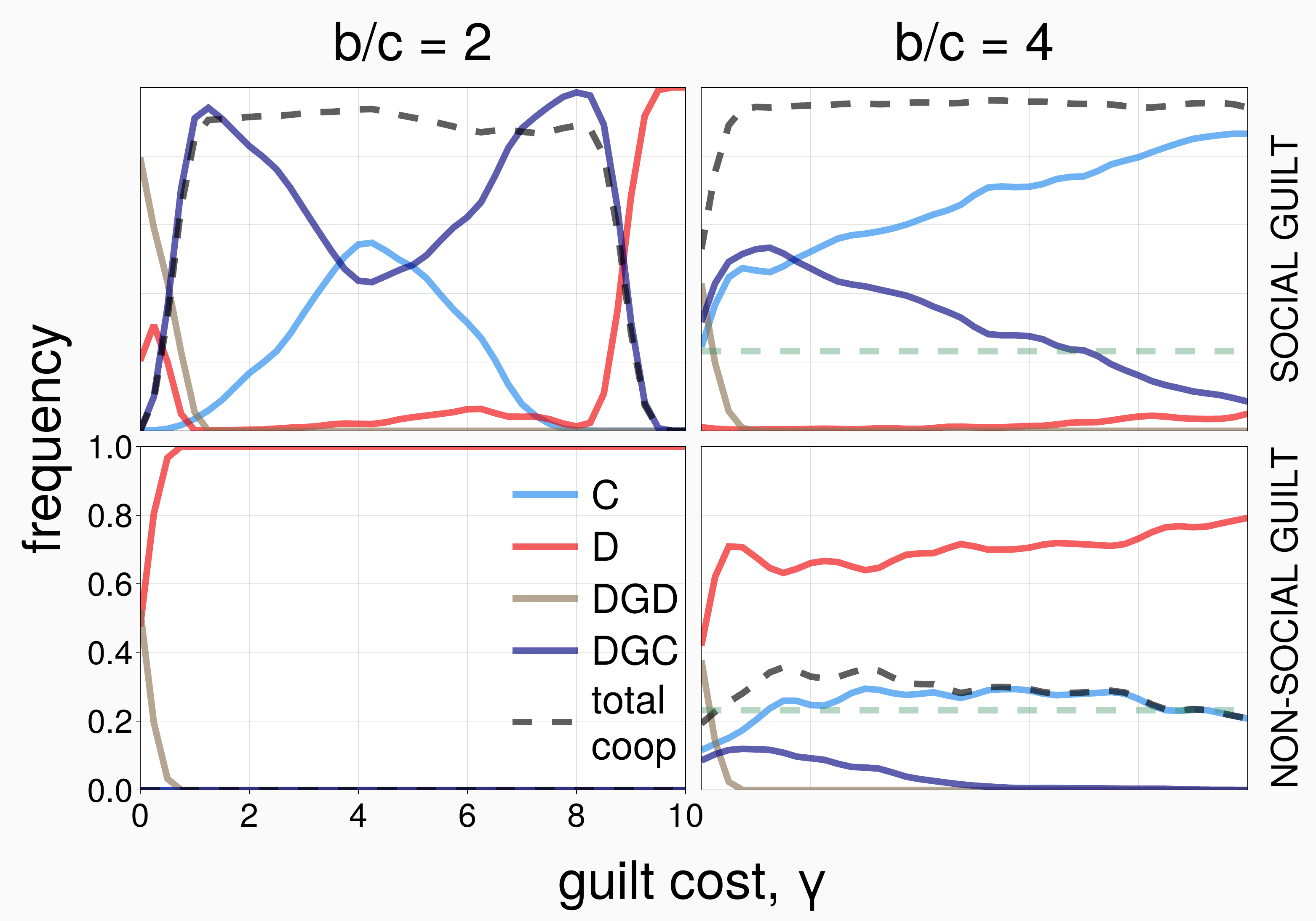}
    \caption{\textbf{Strategies' frequency and the total cooperation level as a function of the guilt cost, $\gamma$, for scale-free networks}. When shown, the dashed green line marks the baseline level of cooperation achieved solely through network reciprocity. Other parameters: population size $N = 1000$, $\Omega = 10$, $\beta = 1.0$. }
    \label{fig:gamma4strats_sf}
    \setlength{\belowcaptionskip}{-10pt}
\end{figure*}

\subsection{Non-social guilt} 

An agent's ongoing guilt level $g$ increases whenever the agent defects, regardless of what the co-player does. 
The payoff matrix for the four strategies C, D,  DGDN, and DGCN, can be written as follows 
\begin{equation} 
 \label{payoff_matrix1} 
\bordermatrix{~ & C & D &  DGDN & DGCN \cr
                  C &R & S &  S & \frac{S+R \Theta}{\Omega }   \cr
                  D & T & P &  P & \frac{P+T \Theta}{\Omega } \cr
                  DGDN & T-\gamma  & P-\gamma  &  P-\gamma  &\frac{P+T \Theta}{\Omega }  -\gamma\cr
                  DGCN  &  \frac{T-\gamma +R \Theta}{\Omega } & \frac{P-\gamma +S \Theta}{\Omega } &  \frac{P-\gamma +S \Theta}{\Omega } & \frac{P-\gamma +R \Theta}{\Omega }
 \cr
                  },
\end{equation}
where we employ $\Theta = \Omega -1$ just for the purpose of neater representation. %Note that the actions C and CGC are essentially equivalent; both considered for the sake of completeness of the strategies set.

\subsection{Social guilt}  
A guilt prone agent feels guilty when defecting only if the co-player acted pro-socially or was observed to feel guilty after defection, viz. through exercising self-punishment or apologising. Thus, in this second model, guilt has a particular social aspect which is missing in the first model.
Now, the payoff matrix is rewritten as follows: 
\begin{equation} 
 \label{payoff_matrix2} 
\bordermatrix{~ & C & D &  DGDS & DGCS \cr
                  C &R & S &  S & \frac{S+R \Theta}{\Omega }   \cr
                  D & T & P &  P & P \cr
                  DGDS & T-\gamma  & P &  P-\gamma  & \frac{P+T \Theta}{\Omega }  -\gamma\cr
                  DGCS  &  \frac{T-\gamma +R \Theta}{\Omega } & P &  \frac{P-\gamma +S \Theta}{\Omega } & \frac{P-\gamma +R \Theta}{\Omega }
 \cr
                  }.
\end{equation}

Notice the differences in the payoff matrices for the interactions between the emotional strategies that defect, i.e. DGD and DGC,  and the unemotional defector D.

 %Note also that these payoff matrices are conceptually different from those used in the situation where commitment and costly apology are used (see \citep{martinez2015apology}):  in the current work,  apology does not  induce a benefit for the co-player.

%With these models, in the next section we aim to answer with our analysis the following questions: 1) Can emotional/guilt behaviours evolve without an ability to express/signal emotion? 2) When can signalling  evolve (upper threshold of signaling cost) and enable guilt  to evolve? \tb{Optional: 3) Can we replace costly signalling with costly observation? Have a balance of both? I suppose costly observation only makes sense if fake signalling is an option, or if signalling is too costly or impossible, or dependant on how many fakers there are around. We can leave these questions for the future work mention, and just keep signalling always trustful and no need for observation costs.}

\subsection{Other results}

For well-mixed populations, please see Figure \ref{fig:dependgamma-appendix}. For square lattice populations, please see Figure \ref{fig:gamma4strats_lattice}. For Scale-Free (SF) networks, please see Figure \ref{fig:gamma4strats_sf}. 

%Furthermore, we have studied several typical simulation runs (for the six strategy model in the main text) to supplement those presented in Fig. \ref{fig:clustering}. We considered several sets of parameters' values, to support the analysis in the main text. Due to size and space limitations, please contact the first author if you would like copies of these figures.

\begin{figure}[t]
    \centering
    \includegraphics[width=\textwidth]{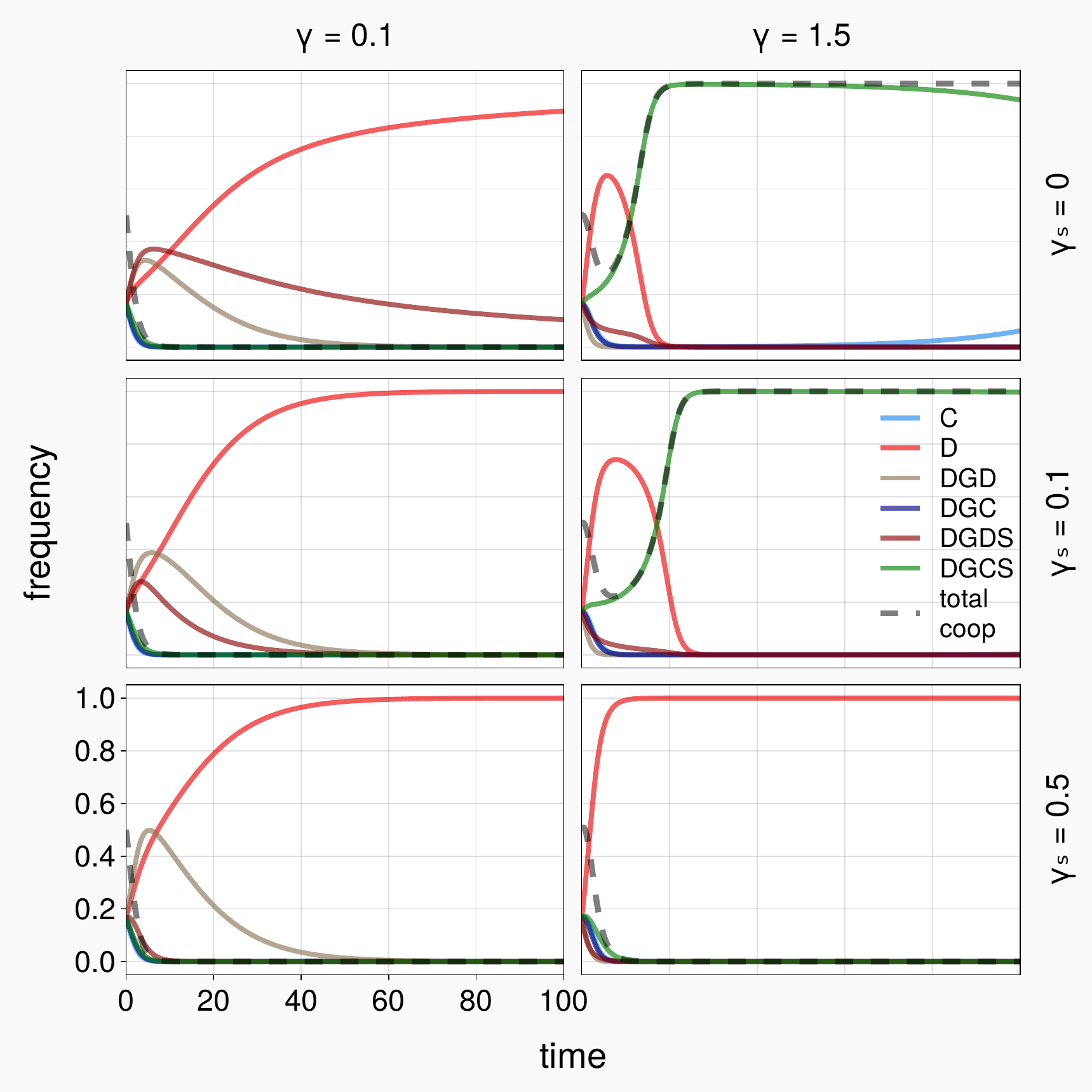}
    \caption{
        \textbf{Time series of strategy frequencies under replicator dynamics for the same parameter settings as in Figure~2 of the main text.} Each trajectory begins from random initial conditions with equal frequency for all six strategies. This figure complements the stationary distribution analysis of Figure~2 by visualizing the transient dynamics and frequency changes over time for different values of $\gamma$ and $\gamma_s$. Other parameters: $\Omega = 10$, $R = 1$, $S = -1$, $T = 2$, and $P = 0$. 
    }
    \label{fig:replicator_dynamics_time_series}
\end{figure}

Figure \ref{fig:replicator_dynamics_time_series} presents time series of the evolutionary dynamics of the six strategies under replicator dynamics, using the same parameter configurations as in Figure 2 of the main text. While Figure 2 illustrates the stationary distributions obtained from Markov transition diagrams, this dynamic simulation provides insight into the transient behavior and temporal progression of strategy frequencies. The replicator dynamics were initialized with equal frequencies across all six strategies (random initial conditions).

\subsection{Equilibrium Analysis (6 strategies)}

\begin{figure}[ht]
    \centering
    \includegraphics[width=0.85\textwidth]{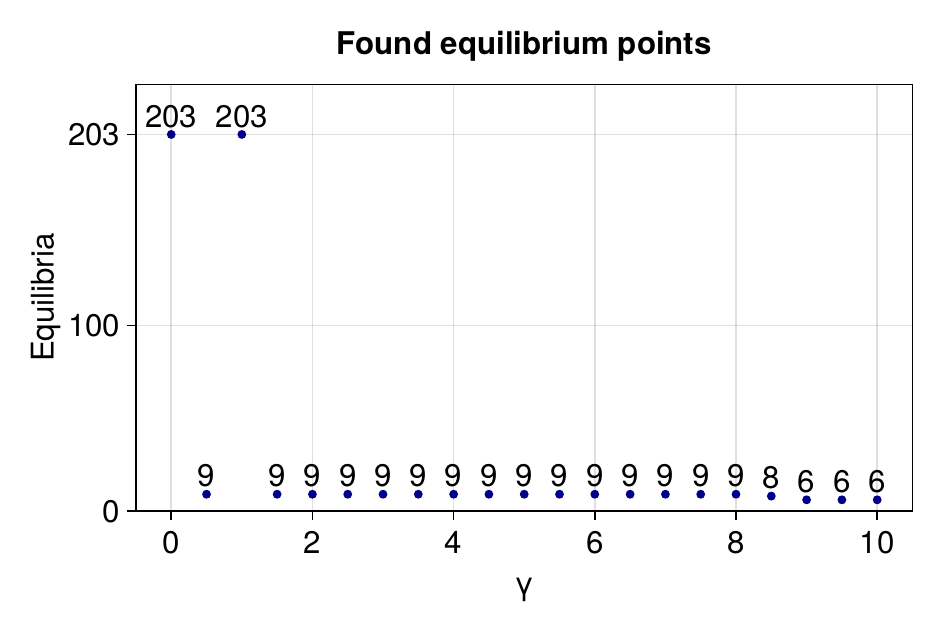}
    \caption{
        \textbf{Total number of equilibria found as a function of the parameter $\gamma$, based on 100{,}000 attempts per value of $\gamma$.} Equilibria were rounded to 3 decimal places and filtered for uniqueness using a Euclidean distance threshold of $10^{-3}$. For low values of $\gamma$, the number of total equilibria appears significantly higher, though this count still includes many near-duplicate or very closely spaced points. Increasing the filtering threshold to group more of these close equilibria would overly suppress valid distinct points at other values of $\gamma$, hence a uniform threshold was maintained. All equilibria found are unstable, except for a small number of stable equilibria observed when $\gamma = 0$. Note that the total number of found equilibria for low $\gamma$ far exceeds the theoretical maximum (203 points for 6 strategies), and we show only the theoretical maximum for the plot, this is due to very close (duplicate) solutions. Other parameters: $\gamma_s = 0.1$, $\Omega = 10$, $R = 1$, $S = -1$, $T = 2$, and $P = 0$.
    }
    \label{fig:total_equilibria}
\end{figure}

\section*{Equilibria Summary}

The total number of equilibria presented in Fig \ref{fig:total_equilibria} demonstrates the inherent complexity of the system under analysis. As shown, the number of equilibria increases significantly at lower values of the parameter \(\gamma\), although this count includes many near-duplicate or closely spaced equilibria, which were filtered using a Euclidean distance threshold of \(10^{-3}\) to ensure uniqueness. Despite the high number of equilibria found, all but a small number of equilibria at \(\gamma = 0\) are unstable. This instability, combined with the large number of equilibria, makes it difficult to gain meaningful insight using this method, as the system's behavior becomes challenging to interpret. The large volume of equilibria, particularly with their proximity to one another and their general instability, complicates the task of identifying distinct and stable solutions, which is essential for understanding the system’s dynamics. 

At $\gamma = 0$, we performed 1,000,000 attempts with the solver to identify stable equilibria, using the same filtering criteria and parameter settings applied in Figure~\ref{fig:total_equilibria}: rounding each solution to three decimal places and removing near-duplicates using a Euclidean distance threshold of $10^{-3}$. Among the results, a small number of stable equilibria were identified. Notably, all of these stable points consist solely of mixtures between the \textbf{D} and \textbf{DGDN} strategies, with no participation from other strategies.

\begin{itemize}
    \item \textbf{Total equilibria found:} 961 (beyond theoretical maximum)
    \item \textbf{Stable equilibria found:} 7
\end{itemize}

\begin{tabular}{@{} l c c c c c c @{}}
\toprule
\textbf{Equilibrium} & \textbf{C} & \textbf{D} & \textbf{DGDN} & \textbf{DGCN} & \textbf{DGDS} & \textbf{DGCS} \\
\midrule
Stable 1 & 0.0 & 0.538587 & 0.461413 & 0.0 & 0.0 & 0.0 \\
Stable 2 & 0.0 & 0.610009 & 0.389991 & 0.0 & 0.0 & 0.0 \\
Stable 3 & 0.0 & 0.560917 & 0.439083 & 0.0 & 0.0 & 0.0 \\
Stable 4 & 0.0 & 0.574551 & 0.425449 & 0.0 & 0.0 & 0.0 \\
Stable 5 & 0.0 & 0.440507 & 0.559493 & 0.0 & 0.0 & 0.0 \\
Stable 6 & 0.0 & 0.464016 & 0.535984 & 0.0 & 0.0 & 0.0 \\
Stable 7 & 0.0 & 0.573832 & 0.426167 & 0.0 & 0.0 & 0.0 \\
\bottomrule
\end{tabular}

%\bibliography{sample}% Produces the bibliography via BibTeX.